\documentclass[twocolumn]{aastex631}

\usepackage{graphicx}
\usepackage{amsmath}
\usepackage[utf8x]{inputenc}
\usepackage[encapsulated]{CJK}
\usepackage{ucs}
\graphicspath{{./}{figs/}}
\usepackage{chngcntr}

\newcommand{\parrot}{\texttt{parrot}}
\newcommand{\eazy}{\texttt{EAzY}}
\newcommand{\eazysfhz}{\eazy-\texttt{sfhz}}
\newcommand{\eazyfsps}{\eazy-\texttt{fsps}}
\newcommand{\prospector}{\texttt{Prospector}}

\newcommand{\msun}{{\rm M}_{\odot}}

\newcommand{\zspec}{z_{\rm spec}}

\shorttitle{The UNCOVER Stellar Population Catalog}
\shortauthors{Wang et al.}

\begin{document}

\title{The UNCOVER Survey: A First-look HST+JWST Catalog of Galaxy Redshifts and Stellar Population Properties Spanning $0.2 \lesssim z \lesssim 15$}

\correspondingauthor{Bingjie Wang}
\email{bwang@psu.edu}

\author[0000-0001-9269-5046]{Bingjie Wang (\begin{CJK*}{UTF8}{gbsn}王冰洁\ignorespacesafterend\end{CJK*})}
\affiliation{Department of Astronomy \& Astrophysics, The Pennsylvania State University, University Park, PA 16802, USA}
\affiliation{Institute for Computational \& Data Sciences, The Pennsylvania State University, University Park, PA 16802, USA}
\affiliation{Institute for Gravitation and the Cosmos, The Pennsylvania State University, University Park, PA 16802, USA}

\author[0000-0001-6755-1315]{Joel Leja}
\affiliation{Department of Astronomy \& Astrophysics, The Pennsylvania State University, University Park, PA 16802, USA}
\affiliation{Institute for Computational \& Data Sciences, The Pennsylvania State University, University Park, PA 16802, USA}
\affiliation{Institute for Gravitation and the Cosmos, The Pennsylvania State University, University Park, PA 16802, USA}

\author[0000-0002-2057-5376]{Ivo Labb\'e}
\affiliation{Centre for Astrophysics and Supercomputing, Swinburne University of Technology, Melbourne, VIC 3122, Australia}
\author[0000-0001-5063-8254]{Rachel Bezanson}
\affiliation{Department of Physics \& Astronomy and PITT PACC, University of Pittsburgh, Pittsburgh, PA 15260, USA}
\author[0000-0001-7160-3632]{Katherine E. Whitaker}
\affiliation{Department of Astronomy, University of Massachusetts, Amherst, MA 01003, USA}
\affiliation{Cosmic Dawn Center (DAWN), Niels Bohr Institute, University of Copenhagen, Jagtvej 128, K{\o}benhavn N, DK-2200, Denmark}

\author[0000-0003-2680-005X]{Gabriel Brammer}
\affiliation{Cosmic Dawn Center (DAWN), Niels Bohr Institute, University of Copenhagen, Jagtvej 128, K{\o}benhavn N, DK-2200, Denmark}
\author[0000-0001-6278-032X]{Lukas J. Furtak}
\affiliation{Department of Physics, Ben-Gurion University of the Negev, P.O. Box 653, Beer-Sheva 8410501, Israel}
\author[0000-0003-1614-196X]{John R. Weaver}
\affiliation{Department of Astronomy, University of Massachusetts, Amherst, MA 01003, USA}
\author[0000-0002-0108-4176]{Sedona H. Price}
\affiliation{Department of Physics \& Astronomy and PITT PACC, University of Pittsburgh, Pittsburgh, PA 15260, USA}
\author[0000-0002-0350-4488]{Adi Zitrin}
\affiliation{Department of Physics, Ben-Gurion University of the Negev, P.O. Box 653, Beer-Sheva 8410501, Israel}

\author[0000-0002-7570-0824]{Hakim Atek}
\affiliation{Institut d'Astrophysique de Paris, CNRS, Sorbonne Universit\'e, 98bis Boulevard Arago, 75014, Paris, France}
\author[0000-0001-7410-7669]{Dan Coe}
\affiliation{Space Telescope Science Institute, Baltimore, MD 21218, USA}
\affiliation{Association of Universities for Research in Astronomy, Inc. for the European Space Agency}
\affiliation{Center for Astrophysical Sciences, Department of Physics \& Astronomy, Johns Hopkins University, Baltimore, MD 21218, USA}
\author[0000-0002-7031-2865]{Sam E. Cutler}
\affiliation{Department of Astronomy, University of Massachusetts, Amherst, MA 01003, USA}
\author[0000-0001-8460-1564]{Pratika Dayal}
\affiliation{Kapteyn Astronomical Institute, University of Groningen, P.O. Box 800, 9700 AV Groningen, The Netherlands}
\author[0000-0002-8282-9888]{Pieter van Dokkum}
\affiliation{Department of Astronomy, Yale University, New Haven, CT 06511, USA}
\author[0000-0002-1109-1919]{Robert Feldmann}
\affiliation{Institute for Computational Science, University of Zurich, Winterhurerstrasse 190, CH-8006 Zurich, Switzerland}
\author[0000-0001-9002-3502]{Danilo Marchesini}
\affiliation{Department of Physics \& Astronomy, Tufts University, Medford, MA 02155, USA}
\author[0000-0002-8871-3026]{Marijn Franx}
\affiliation{Leiden Observatory, Leiden University, P.O.Box 9513, NL-2300 AA Leiden, The Netherlands}
\author[0000-0003-4264-3381]{Natascha F\"{o}rster Schreiber}
\affiliation{Max-Planck-Institut f{\"u}r extraterrestrische Physik, Gie{\ss}enbachstra{\ss}e 1, 85748 Garching, Germany}
\author[0000-0001-7201-5066]{Seiji Fujimoto}
\altaffiliation{Hubble Fellow}
\affiliation{Department of Astronomy, The University of Texas at Austin, Austin, TX 78712, USA}
\author[0000-0002-7007-9725]{Marla Geha}
\affiliation{Department of Astronomy, Yale University, New Haven, CT 06511, USA}
\author[0000-0002-3254-9044]{Karl Glazebrook}
\affiliation{Centre for Astrophysics and Supercomputing, Swinburne University of Technology, Melbourne, VIC 3122, Australia}
\author[0000-0002-2380-9801]{Anna de Graaff}
\affiliation{Max-Planck-Institut f{\"u}r Astronomie, K{\"o}nigstuhl 17, D-69117, Heidelberg, Germany}
\author[0000-0002-5612-3427]{Jenny E. Greene}
\affiliation{Department of Astrophysical Sciences, Princeton University, Princeton, NJ 08544, USA}
\author[0000-0002-0000-2394]{St\'ephanie Juneau}
\affiliation{NSF’s National Optical-Infrared Astronomy Research Laboratory, 950 N. Cherry Avenue, Tucson, AZ 85719, USA}
\author[0000-0002-3838-8093]{Susan Kassin}
\affiliation{Space Telescope Science Institute, Baltimore, MD 21218, USA}
\author[0000-0002-7613-9872]{Mariska Kriek}
\affiliation{Leiden Observatory, Leiden University, P.O.Box 9513, NL-2300 AA Leiden, The Netherlands}
\author[0000-0002-3475-7648]{Gourav Khullar}
\affiliation{Department of Physics \& Astronomy and PITT PACC, University of Pittsburgh, Pittsburgh, PA 15260, USA}
\author[0000-0003-0695-4414]{Michael Maseda}
\affiliation{Department of Astronomy, University of Wisconsin-Madison, Madison, WI 53706, USA}
\author[0000-0002-8530-9765]{Lamiya A. Mowla}
\affiliation{Dunlap Institute for Astronomy \& Astrophysics, Toronto, Ontario, M5S 3H4, Canada}
\author[0000-0002-9330-9108]{Adam Muzzin}
\affiliation{Department of Physics \& Astronomy, York University, Toronto, Ontario, ON MJ3 1P3, Canada}
\author[0000-0003-2804-0648]{Themiya Nanayakkara}
\affiliation{Centre for Astrophysics and Supercomputing, Swinburne University of Technology, Melbourne, VIC 3122, Australia}
\author[0000-0002-7524-374X]{Erica J. Nelson}
\affiliation{Department for Astrophysical \& Planetary Science, University of Colorado, Boulder, CO 80309, USA}
\author[0000-0001-5851-6649]{Pascal A. Oesch}
\affiliation{Department of Astronomy, University of Geneva, Chemin Pegasi 51, 1290 Versoix, Switzerland}
\affiliation{Cosmic Dawn Center (DAWN), Niels Bohr Institute, University of Copenhagen, Jagtvej 128, K{\o}benhavn N, DK-2200, Denmark}
\author[0000-0003-4196-0617]{Camilla Pacifici}
\affiliation{Space Telescope Science Institute, Baltimore, MD 21218, USA}
\author[0000-0002-9651-5716]{Richard Pan}
\affiliation{Department of Physics \& Astronomy, Tufts University, Medford, MA 02155, USA}
\author[0000-0001-7503-8482]{Casey Papovich}
\affiliation{Department of Physics \& Astronomy, Texas A\&M University, College Station, TX, 77843-4242 USA}
\affiliation{George P. and Cynthia Woods Mitchell Institute for Fundamental Physics and Astronomy, Texas A\&M University, College Station, TX, 77843-4242 USA}
\author[0000-0003-4075-7393]{David J. Setton}
\affiliation{Department of Physics \& Astronomy and PITT PACC, University of Pittsburgh, Pittsburgh, PA 15260, USA}
\author[0000-0003-3509-4855]{Alice E. Shapley}
\affiliation{Department of Physics \& Astronomy, University of California: Los Angeles, Los Angeles, CA 90095, USA}
\author[0000-0001-8034-7802]{Renske Smit}
\affiliation{Astrophysics Research Institute, Liverpool John Moores University, 146 Brownlow Hill, Liverpool L3 5RF, UK}
\author[0000-0001-7768-5309]{Mauro Stefanon}
\affiliation{Departament d'Astronomia i Astrofisica, Universitat de Valencia, C. Dr. Moliner 50, E-46100 Burjassot, Valencia, Spain}
\affiliation{Unidad Asociada CSIC ``Grupo de Astrofisica Extragalactica y Cosmologi'' (Instituto de Fisica de Cantabria - Universitat de Valencia)}
\author[0000-0002-1714-1905]{Katherine A. Suess}
\altaffiliation{Hubble Fellow}
\affiliation{Kavli Institute for Particle Astrophysics and Cosmology and Department of Physics, Stanford University, Stanford, CA 94305, USA}
\author[0000-0002-5522-9107]{Edward N. Taylor}
\affiliation{Centre for Astrophysics and Supercomputing, Swinburne University of Technology, Melbourne, VIC 3122, Australia}
\author[0000-0003-2919-7495]{Christina C. Williams}
\affiliation{NSF's National Optical-Infrared Astronomy Research Laboratory, Tucson, AZ 85719, USA}
\affiliation{Steward Observatory, University of Arizona, Tucson, AZ 85721, USA}

\begin{abstract}
The recent UNCOVER survey with the James Webb Space Telescope (JWST) exploits the nearby cluster Abell 2744 to create the deepest view of our universe to date by leveraging strong gravitational lensing. In this work, we perform photometric fitting of more than 50,000 robustly detected sources out to $z \sim 15$. We show the redshift evolution of stellar ages, star formation rates, and rest-frame colors across the full range of $0.2 \lesssim z \lesssim 15$. The galaxy properties are inferred using the \texttt{Prospector} Bayesian inference framework using informative \texttt{Prospector}-$\beta$ priors on masses and star formation histories to produce joint redshift and stellar population posteriors, and additionally lensing magnification is performed on-the-fly to ensure consistency with the scale-dependent priors. We show that this approach produces excellent photometric redshifts with $\sigma_{\rm NMAD} \sim 0.03$, of a similar quality to the established photometric redshift code \texttt{EAzY}. In line with the open-source scientific objective of the Treasury survey, we publicly release the stellar population catalog with this paper, derived from the photometric catalog adapting aperture sizes based on source profiles. This release~\footnote{The catalog and all related documentation are accessible via the UNCOVER survey webpage: \url{https://jwst-uncover.github.io/DR2.html\#SPSCatalogs} with a copy deposited to Zenodo: \dataset[doi:10.5281/zenodo.8401181]{https://doi.org/10.5281/zenodo.8401181}} includes posterior moments, maximum-likelihood spectra, star-formation histories, and full posterior distributions, offering a rich data set to explore the processes governing galaxy formation and evolution over a parameter space now accessible by JWST.
\end{abstract}


\section{Introduction}

The Hubble Space Telescope (HST) has secured a lasting legacy in mapping the key epochs of galaxy assembly via Treasury programs (e.g., \citealt{2011ApJS..197...35G,2012ApJ...758L..17B}). JWST, with its sensitive NIRCam imaging and NIRSpec spectroscopy at 1--5 $\mu$m \citep{Boker2023,Rieke2023,Rigby2023}, promises to not only extend this legacy, but also reveal new mysteries of the distant universe out to $z \sim 15$ and beyond. Thus far, data from the JWST Early Release Science programs have already shed new light on the early phases of galactic evolution (e.g., \citealt{Treu2022,2022arXiv221105792F}). The JWST extragalactic Treasury survey, Ultradeep NIRSpec and NIRCam ObserVations before the Epoch of Reionization (UNCOVER; \citealt{Bezanson2022}), has completed its primary NIRCam imaging observation on and around the gravitational lensing cluster Abell~2744 at $z=0.308$ in November 2022. The images reach depths of $\sim30$ AB magnitudes. After accounting for gravitational lensing, the intrinsic depths reach $\gtrsim31-32$ AB magnitudes, making UNCOVER the deepest survey in Cycle~1 of JWST observations. The quest for a coherent understanding of the universe in the newly observed parameter space is just beginning.

For the first time, it is possible to observe galaxies and candidates spanning $0.2 \lesssim z \lesssim 15$ and infer their stellar population properties. 
In the context of UNCOVER, spectroscopically confirmed galaxies at $z>12$ and an X-ray luminous supermassive black holes at $z=10.1$ are studied in detail in \citet{wang2023:z12} and \citet{goulding2023}, respectively, whereas a systematic search for $\zspec \gtrsim 9$ sources is carried out in \citet{Fujimoto2023}. The core of this paper is to present a galaxy catalog containing key stellar population metrics over the full dynamic range probed by the survey. This work is part of the second Data Release (DR2) from UNCOVER. In accordance with the open-data intent of the Treasury survey, we have made publicly available the imaging mosaics \citep{Bezanson2022}, updated strong lensing model \citep{Furtak2022}, and first-look photometric catalogs \citep{Weaver2023}. This paper constitutes the final installation of the UNCOVER DR2: inferred galaxy properties, including redshifts, stellar masses, metallicities, ages, star formation rates (SFRs), dust attenuation values, and fractional mid-infrared active galactic nuclei (AGN) luminosities. It is accompanied by the magnification factor, and additionally the radial magnification, tangential magnification, and shear, consistent with the inferred redshifts. The full posterior distributions are also released.

The main data products in this work are inferred using the \prospector\ Bayesian framework \citep{Johnson2021}, with two notable modifications. First, we optimize our priors for recovering accurate photometric redshifts by including observationally-motivated, joint priors on stellar mass, stellar metallicity, and star formation history (SFH) from \prospector-$\beta$ \citep{Wang2023}.
 Second, we solve the magnification--redshift relationship on-the-fly within \prospector\ to take advantage of the mass-dependent priors, in contrast to the traditional approach where physical parameters are scaled by magnification factors post-fit.

Inferring redshifts across a wide range of distances and galaxy properties based solely on photometric data pushes the stellar population models into new, exciting, and largely uncalibrated regimes. Comparisons between different photo-$z$ codes have clearly highlighted that accuracy in this space depends strongly on the assumptions that go into the code (e.g., \citealt{2023ApJ...942...36K}). Given the novelty of our \prospector\ model, we cross-check our redshifts with those from the established template-fitting code \eazy\ \citep{Brammer2008}. We additionally analyze the effect of varying template sets within \eazy\ by comparing results from the default set \texttt{fsps} to \texttt{sfhz}. The latter incorporates redshift-dependent SFHs and a realistic emission line model at $z \sim 8$ \citep{Carnall2023}.

The structure of this paper is as follows. Section~\ref{sec:data} summarizes the photometric data. Section~\ref{sec:method} details the SED modeling. Section~\ref{sec:res} presents the inferred parameters. Section~\ref{sec:issues} discusses known problems in the photometry and in the modeling, and how these may affect the accuracy of the inferred stellar population parameters. Section~\ref{sec:sum} concludes with a brief summary and the format of our catalogs.

Where applicable, we adopt the best-fit cosmological parameters from the 9 yr results from the Wilkinson Microwave Anisotropy Probe mission: $H_{0}=69.32$ ${\rm km \,s^{-1} \,Mpc^{-1}}$, $\Omega_{M}=0.2865$, and $\Omega_{\Lambda}=0.7135$ \citep{2013ApJS..208...19H}, and a Chabrier initial mass function (IMF; \citealt{Chabrier2003}). Unless otherwise mentioned, we report the median of the posterior, and 1$\sigma$ error bars are the 16th and 84th percentiles.

\section{Data\label{sec:data}}

\begin{figure*}
\gridline{
\fig{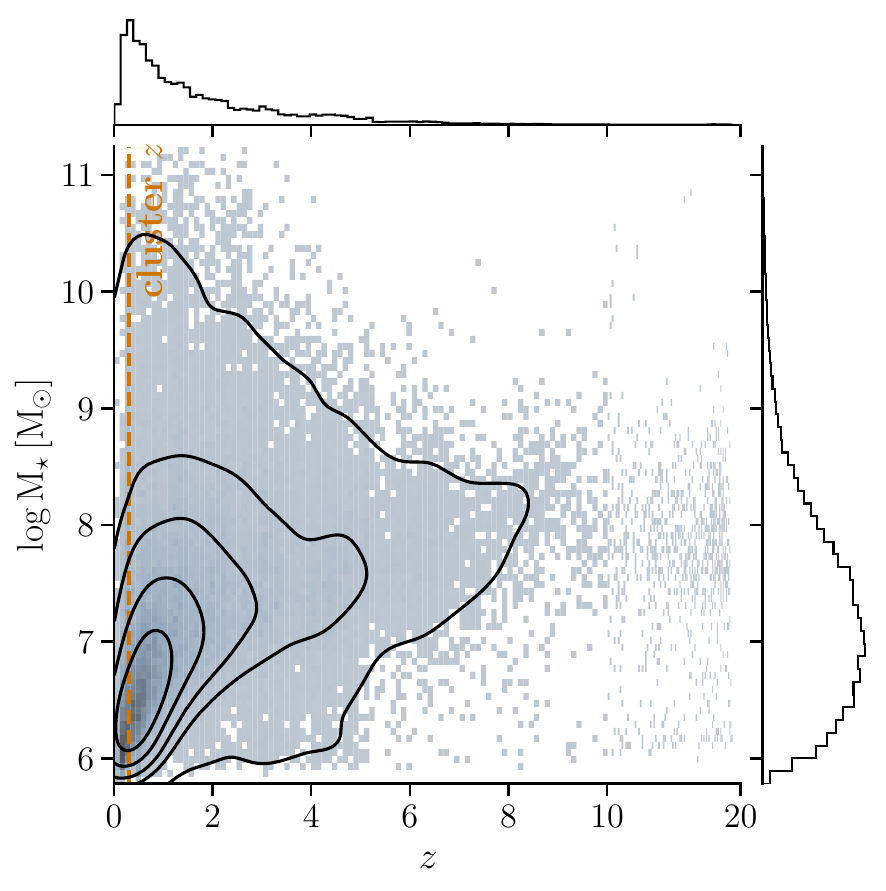}{0.7\textwidth}{}
}
\gridline{
\fig{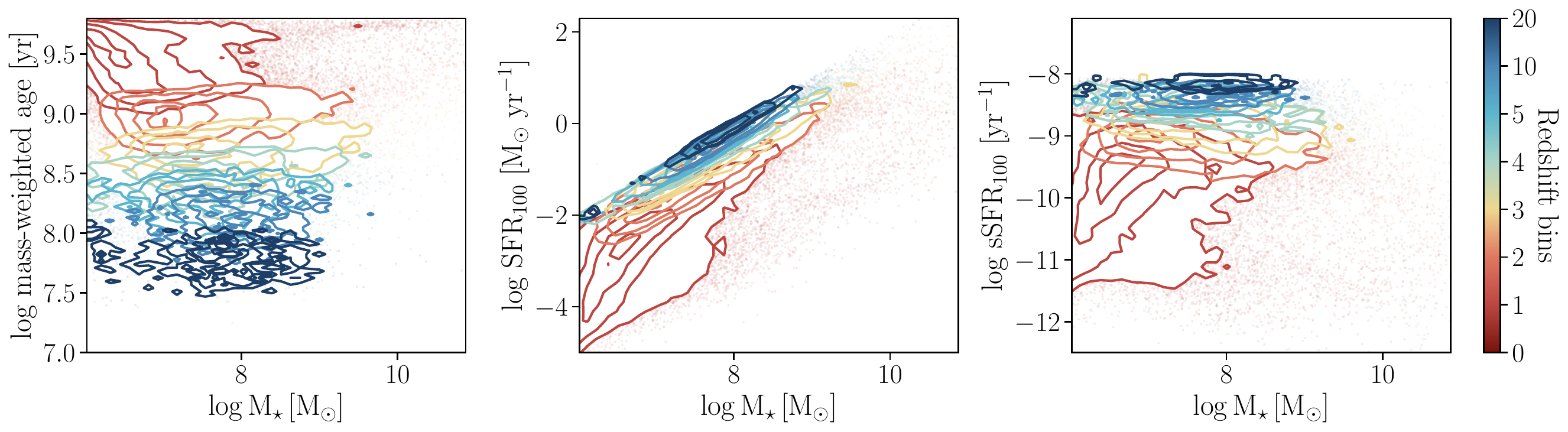}{0.99\textwidth}{}
}
\caption{An overview of the key inferred parameters in our catalog. (upper panel) Joint distribution of redshifts and stellar masses. The x-axis is in linear scale at $z \leq 10$, while in logarithmic scale at $z>10$. The redshift at which the Abell~2744 cluster resides ($z=0.308$) is indicated by an orange dashed line. The secondary peak in the mass distribution at $\sim 10^6 \, \msun$ is mainly due to the large population of globular clusters. (lower panel) Mass-weighted ages, star formation rates (100 Myr), and specific star formation rates (100 Myr) as functions of stellar masses in redshift bins.
\label{fig:overview}}
\end{figure*}

The photometry includes all public JWST/NIRCam, HST/ACS and HST/WFC3 imaging of Abell~2744 available to date. Specifically, the JWST data includes the Cycle 1 Treasury program UNCOVER covering $\sim$~45 arcmin$^2$ (PIs Labb\'{e} \& Bezanson, JWST-GO-2561; \citealt{Bezanson2022}), the Early Release Science program GLASS (PI: Treu, JWST-ERS-1324; \citealt{Treu2022}), and a Director's Discretionary program (JWST-DD-2756, PI: Chen). These observations span $\sim 1-5 \mu$m in observed frame in 8 filters: F090W, F115W, F150W, F200W, F277W, F356W, F410M, and F444W. The HST data, taken from the public archive, consist of HST-GO-11689 (PI: Dupke), HST-GO-13386 (PI: Rodney), HST-DD-13495 (PI: Lotz; \citealt{Lotz2017}), and HST-GO-15117 (PI: Steinhardt; \citealt{Steinhardt2020}). These additional observations span $\sim 0.4-1.6\,\mu$m in observed frame in 7 filters: F435W, F606W, F814W, F105W, F125W, F140W, and F160W. Details of these programs including imaging depths, and ancillary information are summarized in Tables 1 and 3 in \citet{Bezanson2022}.
A small subsample has detections or upper limits from DUALZ (Deep UNCOVER-ALMA Legacy High-Z Survey; \citealt{Fujimoto2023:alma}). The transmission curve of this ALMA band is approximated as a top-hat function spanning 1249.9 to 1351.0 $\mu$m.

As part of UNCOVER DR2, we have released F277W+F356W+F444W-selected photometric catalogs containing total fluxes for 61,648 sources \citep{Weaver2023}.
In this work we fit galaxy SEDs to the ``super-catalog," which uses adaptive aperture selection based on their isophotal areas following \citet{Labbe2003}. We share publicly the inferred parameters for the full sample. However, it is worth noting that a subsample of 55,613 objects is deemed to have reliable photometry, i.e., \texttt{use\_phot}=1 in the photometric catalog. Furthermore, 15,861 objects in this subsample are flagged as possible blends (\texttt{flag\_kron} = 1), and therefore their photometry is not corrected to total based on a kron ellipse but instead is corrected assuming a point-like morphology. Photometry performed on simulated galaxies with realistic sizes indicate that the total photometry of these objects may be underestimated by a factor of 2 in 0.32-0.70\arcsec{} apertures, with larger 1.00-1.40\arcsec{} apertures miss $<10$\% of the total flux. Consequently, physical parameters derived in this work including stellar mass and rest-frame fluxes are liable to be underestimated. Details on the selection criteria for the \texttt{use\_phot} and \texttt{flag\_kron} flags are presented in Section 4.6 in \citet{Weaver2023}.

A total of $\sim$~400 reliable, unblended sources in the UNCOVER photometric catalogs have spectroscopic redshifts collected from the NASA/IPAC Extragalactic Database, and from the literature. The latter consists of measurements taken with HST as part of the Grism Lens Amplified Survey from Space (GLASS; \citealt{Treu2015}), and with the Multi Unit Spectroscopic Explorer (MUSE; \citealt{Richard2021}).
We utilize this sample to evaluate the accuracy of our photometric redshift recovery.

\section{SED fitting\label{sec:method}}

This section describes the core of this work---inferring physical parameters of the galaxy populations from the photometry. An overview of the inferred parameters of our catalog is shown in Figure~\ref{fig:overview}.

In what follows, we first describe our modified \prospector\ model, which includes physically motivated priors to optimally produce joint constraints of redshifts, stellar masses, and other key stellar population metrics across any redshift range, and a consistent treatment of lensing magnification during model fitting. We then outline the adopted \eazy\ settings, and its two template sets, \texttt{fsps} and \texttt{sfhz}. A minimum error floor of 5\%\ is imposed in all fits to reflect the additional systematic uncertainties---the calibration uncertainties of JWST/NIRCam at this early stage \citep{Boyer2022}, as well as uncertainties in the stellar population synthesis (SPS).

\subsection{Prospector}

\begin{deluxetable*}{p{0.1\textwidth} p{0.4\textwidth} p{0.4\textwidth}}
\tablecaption{\prospector-$\beta$ Parameters and Priors Adopted in This Paper\label{tab:theta}}
\tablehead{
\colhead{Parameter} & \colhead{Description} & \colhead{Prior}
}
\startdata
$z$ & redshift & uniform: $\mathrm{min}=0$, $\mathrm{max}=20$ \\
$\log~(\mathrm{M}/\mathrm{M}_{\odot})$ & total stellar mass formed & mass functions in \citet{Leja2020} \citep{Wang2023}\\
SFH & ratio of SFRs in adjacent log-spaced time bins & SFH($M, z$) \citep{Wang2023}\\
$\log~(\mathrm{Z}^{\star}/\mathrm{Z}_{\odot})$ & stellar metallicity & Gaussian approximating the $\mathrm{M}$--Z$^{\star}$ relationship of \cite{Gallazzi2005} \\
$n$ & power law index for a \citet{Calzetti2000} attenuation curve & uniform: $\mathrm{min}=-1.0$, $\mathrm{max}=0.4$ \\
$\hat{\tau}_{\rm dust, 2}$ & optical depth of diffuse dust \citep{Charlot2000} & truncated normal: $\mathrm{min}=0$, $\mathrm{max}=4$, $\mu=0.3$, $\sigma=1$\\
$\hat{\tau}_{\rm dust, 1} / \hat{\tau}_{\rm dust, 2}$ & ratio between the optical depths of birth cloud dust and diffuse dust \citep{Charlot2000} & truncated normal: $\mathrm{min}=0$, $\mathrm{max}=2$, $\mu=1$, $\sigma=0.3$ \\
$\log~f_{\mathrm{AGN}}$ & ratio between the object's AGN luminosity and its bolometric luminosity & uniform: $\mathrm{min}=-5$, $\mathrm{max}=\log 3$ \\
\hline
$\log~\tau_{\mathrm{AGN}}$ & optical depth of AGN torus dust & uniform: $\mathrm{min}=\log 5$, $\mathrm{max}=\log 150$ \\
$\log~(\mathrm{Z}_{\mathrm{gas}}/\mathrm{Z}_{\odot})$ & gas-phase metallicity & uniform: $\mathrm{min}=-2.0$, $\mathrm{max}=0.5$ \\
$\mathrm{q_{PAH}}$ & fraction of grain mass in PAHs \citep{Draine2007} & truncated normal: $\mathrm{min}=0$, $\mathrm{max}=7$, $\mu=2$, $\sigma=2$\\
\enddata
\end{deluxetable*}

\begin{figure*}
\gridline{
\fig{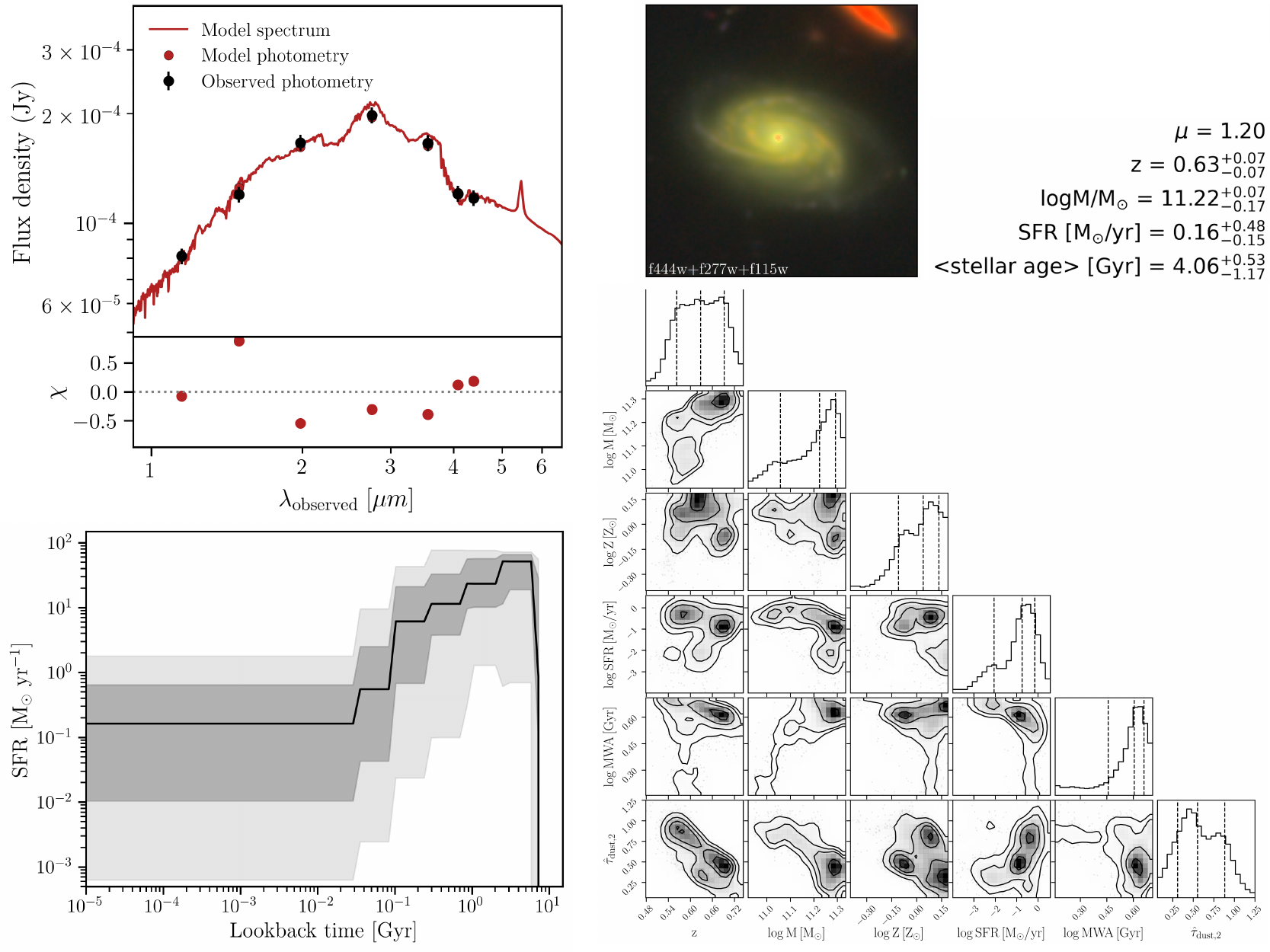}{0.99\textwidth}{}
}
\caption{An example from the UNCOVER field. The upper left shows the SED of the galaxy. The observed photometry is plotted as black dots, while the maximum-likelihood photometry and spectrum are shown in red. The lower left shows the inferred SFH of the galaxy. The x-axis is the lookback time in Gyr, ans the y-axis is the SFR in $\rm M_\odot \, yr^{-1}$. The RGB color composite is made from F444W, F277W, and F115W bands. The lower right is a corner plot showing the marginal and joint posterior distributions of redshift, stellar mass, stellar metallicity, star formation rate, mass-weighted age, and optical depth of the diffuse dust.
\label{fig:sed_ex}}
\end{figure*}

The main data products in the catalog are inferred using the \prospector\ Bayesian inference framework \citep{Johnson2021}. The building blocks of these fits, i.e., the simple stellar populations (SSPs), come from FSPS \citep{Conroy2010}, where we adopt MIST isochrones \citep{Choi2016,Dotter2016} and MILES stellar library \citep{2006MNRAS.371..703S}. The composite stellar populations (CSPs) are modeled with \prospector-$\beta$ \citep{Wang2023}, which follows \prospector-$\alpha$ \citep{Leja2017} in many components. 
In this section, we begin with the common elements in \prospector-$\alpha$, and then proceed to the new additions from \prospector-$\beta$, namely the joint priors on stellar mass, stellar metallicity, and SFH. We end with a novel modification in the likelihood calculation which allows for a self-consistent treatment for lensing magnification during model fitting.

The SFH is modeled as mass formed in 7 logarithmically-spaced time bins, and assumes a continuity prior to ensure smooth transitions between bins \citep{Leja2019}. The scheme for the age bins is refined in this work. For $z<3$, we keep the conventional definition in which the first two bins are always 30 Myr and 100 Myr respectively, whereas for $z \geq 3$, we only require that the first bin is always 13.47 Myr. In both cases, the last bin is 10\% of the age of the universe at a given redshift, and the intervening bins are evenly spaced in logarithmic time. This new scheme ensures that no age bins are overly wide in the early universe. 
Nebular emission is included using a pre-computed \texttt{Cloudy} grid \citep{Byler2017}.
Dust is described using a two-component model \citep{Charlot2000} with a flexible dust attenuation curve \citep{Noll2009}. We also fit for the stellar metallicity, and the normalization and dust optical depth of mid-infrared AGNs. Dust emission is included in all fits \citep{Draine2007}, with the mass fraction of polycyclic aromatic hydrocarbons left free. 
The attenuation of the intergalactic medium (IGM) is assumed to follow \citet{Madau1995}.

In contrast to previous large-scale applications of \prospector\ to broadband photometry, which typically fix redshift to a value determined by an external photo-$z$ code (e.g., \citealt{Leja2019b}), here we fit directly for redshift and use informed priors for the stellar mass, stellar metallicity, and SFH in the \prospector-$\beta$ model. We opt for this approach to take the full advantage of the \prospector\ Bayesian inference framework, and thus to obtain consistent joint constraints on the probability distribution of the full posterior. Following \citet{Wang2023}, we first include a mass prior ${\rm P(logM^{\star}} | z)$, constructed from the observed mass functions between $0.2 < z < 3$ \citep{Leja2020}. For $z < 0.2$ and $z > 3$, we take the nearest-neighbor solution, i.e., the $z$ = 0.2 and $z$ = 3 mass functions. This applies the mass function as a prior where it is well-measured, and avoids relying on simulation predictions while making a reasonable null hypothesis in the absence of observational constraints.
Second, we use a dynamic SFH prior, meaning that the shape of the SFH is dependent on redshift and stellar mass. The expectation value of this prior is matched to the cosmic SFR density \citep{Behroozi2019}; in other words, it encourages rising histories early in the universe, and falling histories late in the universe. This prior additionally reflects the consistent observational finding that massive galaxies form much earlier than low-mass galaxies \citep{Cowie1996,Thomas2005} by introducing a hyper-parameter that scales the expectation value for the shape of SFH with mass, such that massive galaxies have more falling SFHs and low-mass galaxies have more rising SFHs. As noted in \citet{Wang2023}, even though the high-redshift constraints on galaxy evolution remain uncertain, these priors still represent a better expectation then the null expectation of uniform priors.
Third, we place a prior based on the stellar mass--stellar metallicity relationship measured from the SDSS \citep{Gallazzi2005}. Following \citet{Leja2019}, we take the conservative approach of widening the confidence intervals from this relationship by a factor of 2 to account for potential unknown systematics or redshift evolution. This prior neglects predictions from galaxy formation models, which suggest a smaller scatter and a relatively strong redshift evolution \citep{Ma2016,Feldmann2023}.

The complete list of the free parameters and their associated priors is summarized in Table~\ref{tab:theta}. Only the first 8 parameters in the table are used to infer physical parameters that are provided in the first public release. The other parameters are used as nuisance parameters in the great majority of cases, and will be examined further in the future.

It is worth noting that we correct for lensing magnification simultaneously while fitting the full set of parameters. Gravitational lensing is in general achromatic (i.e., the deflection angle of a light ray is independent of its wavelength), meaning that colors are conserved. Therefore, while the magnification factor, $\mu$, depends on redshift and source position, it is not conventionally accounted for in the process of SED fitting. Rather, the modeled fluxes and the scale-dependent physical parameters such as stellar mass are often divided by $\mu$ post-fit, or the observed fluxes are demagnified before fitting using external redshift information. However, scale-{\textit{dependent}} priors in \prospector-$\beta$ necessitate a self-consistent treatment of $\mu$. We devise a simple method: given a source position, we read in the convergence, $\kappa$, and shear, $\gamma$, from the 0.1\arcsec/pixel resolution maps provided as part of UNCOVER DR \citep{Furtak2022}~\footnote{The version used in this work is \texttt{v1.1}, which incorporates new observational constraints bringing the lens plane image reproduction root-mean-square (RMS) of the model down to $\Delta_{\rm{RMS}} = 0.51\arcsec$.}. These maps are normalized such that $D_{\rm ds}/D_{\rm s} = 1$, where $D_{\rm ds}$ is the angular diameter distance between the lens and the source planes, and $D_{\rm s}$ is the distance between the source plane and the observer. Then in each draw of the redshift during sampling, we multiply $\kappa$ and $\gamma$ by the $D_{\rm ds}/D_{\rm s}$ of the source, and magnify the model photometry by $\mu$, which is calculated as
\begin{equation}
	\mu = \frac{1}{ \left| (1-\kappa)^2-\gamma^2 \right|}.
	\label{eq:mu}
\end{equation}
Critically, magnifying the model fluxes and demagnifying the observations are not equivalent due to the definition of likelihood in \prospector:
\begin{equation}
    {\rm ln}\mathcal{L} = \sum_{n=1}^{N} {\rm ln}\left[ \frac{1}{\sqrt{2 \pi \sigma^2}} \cdot {\rm exp}^{ {\frac{\left(x_n - \alpha \right)^2}{-2\sigma^2}} }  \right],
\end{equation}
where $\sigma$ is the observed uncertainties, $x_n$ is the observed flux in the $n$th photometric band, and $\alpha$ is the model uncertainties. The often neglected pre-factor is properly included here because \prospector\ offers the functionality of estimating the observational uncertainties simultaneously with the object parameters. In the context of the magnification calculation, if we change the observational uncertainties with redshift, the pre-factor works to minimize these uncertainties, or in other words to maximize $\mu$. Therefore, we must magnify the model fluxes for the likelihood to behavior correctly.

The posterior space is sampled with the nested sampler \texttt{dynesty} \citep{Speagle2020}, and a neural net emulator, dubbed \texttt{parrot}, which mimics SPS models is used to decrease the runtime \citep{Alsing2020,Mathews2023}. Given that this work is the first large-scale application of the emulator beyond verification tests with the 3D-HST catalogs \citep{Mathews2023}, we test its accuracy compared to full FSPS evaluations in Appendix~\ref{app:is}. We also supply further diagnostics on photometric residuals in Appendix~\ref{app:residual}. An example of an SED fit is shown in Figure~\ref{fig:sed_ex}.

\subsection{EAzY}

\eazy\ is a galaxy photometric redshift code that fits the observed SEDs as a non-negative sum of templates by minimizing the $\chi^2$ statistics \citep{Brammer2008}. \eazy\ offers flexibility in fitting different data sets through the modification of templates. We start by the standard publicly available \texttt{fsps} template set, which consists of 12 templates spanning a range of colors. As pointed out by, e.g., \citet{Steinhardt2022}, the standard method of running \eazy\ can allow for nonphysical contributions from templates that are older than the age of the universe at a given redshift thereby biasing the best fits. The latest \texttt{sfhz} template set is designed to mitigate this problem. Redshift dependence is introduced in the templates so that SFHs starting earlier than the age of the universe at a given epoch are disallowed. A template from the JWST/NIRSpec observation of a $z\sim8$ extreme emission-line galaxy is also added in this set \citep{Carnall2023}, in order to expand the template set to include the more exotic emission-line contributions that have been observed at the highest redshifts.

In consistency with the \eazy\ settings in \citet{Weaver2023}, we turn off the priors on the magnitude and the UV slope.
We do not apply the usual methodology of an iterative application of photometric zero-point offsets either.
The agreement between the spectroscopic and photometric redshifts marginally worsens after applying the correction, though we caution that the redshifts in this field have a strong selection function due to the foreground cluster and the relatively low number of spectroscopic redshifts. We defer a full exploration of the photo-$z$ accuracy in this field to future works which will exploit forthcoming medium-band and grism redshifts.

\section{Inferred Stellar Population Parameters\label{sec:res}}

We now present the parameters contained in our stellar population catalogs, an overview of which is shown in Figure~\ref{fig:overview}. Here we also compare the photometric redshifts inferred using \prospector-$\beta$, \eazysfhz, and \eazyfsps\ as a quality check.

\subsection{Photometric Redshifts\label{subsec:photoz}}

\begin{figure*}
  \centering
    \includegraphics[width=0.99\textwidth]{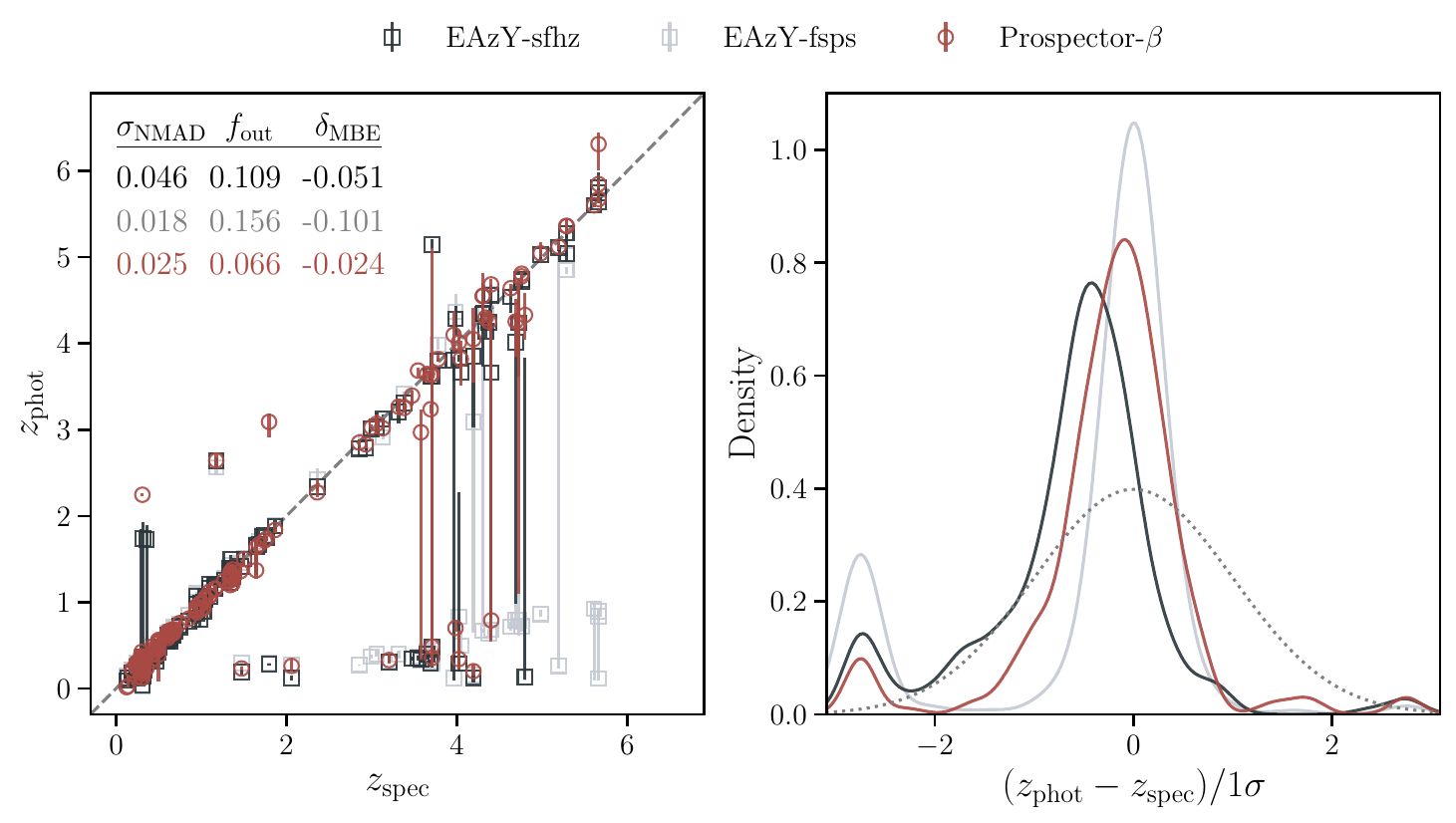}
\caption{Accuracy of the photometric redshifts. (Left) The photometric redshifts inferred from the three different settings described in Section~\ref{sec:method} are plotted against known spectroscopic redshifts. Also included are the summary statistics quantifying the scatter ($\sigma_{\rm NMAD}$; Equation~\ref{eq:nmad}), outlier fraction ($f_{\rm out} (|\Delta z| > 0.15)$), and bias ($\delta_{\rm MBE}$; Equation~\ref{eq:mbe}). These statistics suggest that we can achieve high-quality redshifts for bright objects. (Right) The distributions of photometric redshift residuals normalized by the $1\sigma$ width of the posteriors are illustrated by kernel density estimations. A unit Gaussian is overplotted as a gray dashed curve to guide the eye. The data are clipped to be within $3\sigma$ of the unit Gaussian. This calibration plot suggests that all three codes tend to overestimate the uncertainties for typical objects, and underestimate the uncertainties for the outliers.
\label{fig:zspec}}
\end{figure*}

\begin{figure*}
\gridline{
\fig{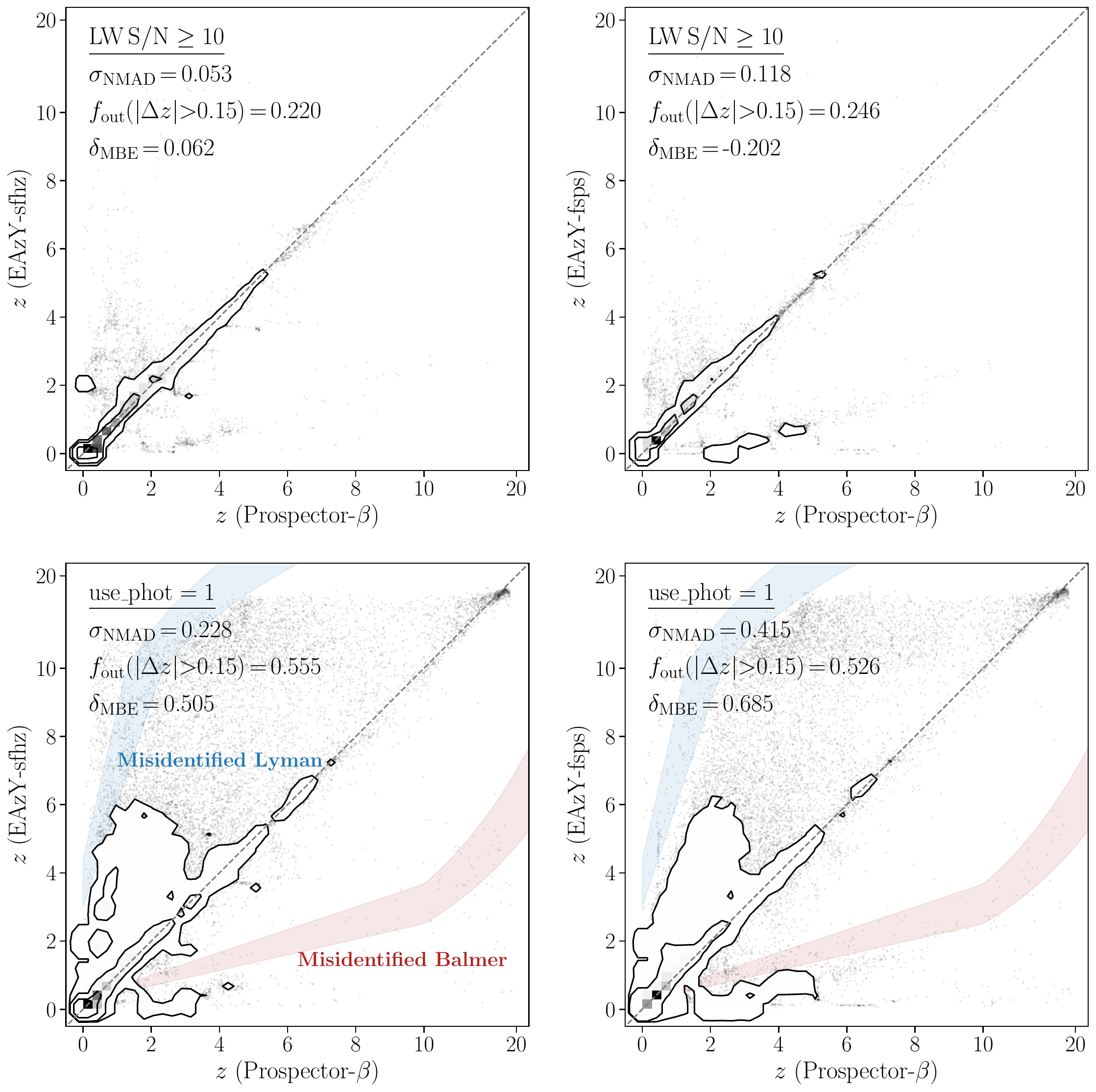}{0.99\textwidth}{}
}
\caption{Comparison between \eazy\ and \prospector\ redshifts. (Upper panel) Photometric redshifts inferred from \prospector-$\beta$ are plotted against those fitted using the latest template set \eazysfhz\ and the standard template set \eazyfsps, respectively. Here only the subset satisfying S/N $\geq$ 10 in the long-wavelength detection bands (F277W, F356W, and F444W) is shown. All the statistics are computed by taking the \prospector-$\beta$ redshifts as ``truth'' for the ease of comparison. The scales on both axes are switched from linear to logarithmic at $z = 10$. (Lower panel) Similar to the upper panel, but the photometric redshifts from the full set of sources with reliable photometry are included (i.e., \texttt{use\_phot}=1). Break confusions (a combination of Balmer, 4000\r{A}, Lyman-$\alpha$, and Lyman-continuum) are indicated by the colored shades. \label{fig:eazy_prosp}}
\end{figure*}

We show the comparison between our recovered photometric redshifts and available spectroscopic redshifts in Figure~\ref{fig:zspec}.
The scatter in residuals is quantified using the normalized median absolute deviation (NMAD; \citealt{Hoaglin1983}) given its advantage of being less sensitive to outliers than standard indicators, e.g., RMS. It is defined as
\begin{equation}
	\sigma_{\rm NMAD} = 1.48 \times {\rm median |\Delta z|},
 	\label{eq:nmad}
\end{equation}
where $\Delta z = (z_{\rm phot} - z_{\rm spec}) / (1+z_{\rm spec})$. We additionally quantify an outlier fraction, $f_{\rm out}$, in which we define a catastrophic outlier as one with $|\Delta z| > 0.15$, and bias calculated using the mean bias error (MBE) as
\begin{equation}
	\delta_{\rm MBE} = \frac{1}{n} \sum_{i=1}^{n} \Delta z.
	\label{eq:mbe}
\end{equation}

Overall, \prospector-$\beta$ performs favorably. \eazy{}'s performance depends on the adopted template set: the \texttt{fsps} template set has a low scatter but a high outlier fraction, while the \texttt{sfhz} template has middle-of-the-road performance in both. This template-dependent performance from \eazy\ is liable to change further as the photometric extraction is improved, and more generally template-based fitting appears to be more sensitive to the choices made during photometric catalog construction. Importantly, as the spectroscopic sample unavoidably has a strong selection bias, we caution not to over-interpolate these results. 

We proceed to compare the photometric redshifts from \prospector-$\beta$, \eazysfhz, and \eazyfsps\ in Figure~\ref{fig:eazy_prosp}. For the ease of comparing the statistics, we take results from \prospector-$\beta$ as the reference point; that is, $\Delta z = (z_{\rm EAzY} - z_{\rm Prospector-\beta}) / (1+z_{\rm Prospector-\beta})$.
\eazysfhz\ notably produces smaller scatter and bias than \eazyfsps\ when comparing to the \prospector-$\beta$ redshifts, although both agree reasonably well on a high signal-to-noise (S/N) subsample. The agreement deteriorates, however, when all the reliable photometric data is included. As seen in the lower panel of Figure~\ref{fig:eazy_prosp}, break confusion is not the main cause. Combining with the findings from the spectroscopic sample, model mismatch is a more likely explanation. The scheduled spectroscopic observation of UNCOVER will allow for a more definitive conclusion.

The existence of $z \gtrsim 9$ sources in this catalog is apparent in both Figures~\ref{fig:overview} and \ref{fig:eazy_prosp}. Possible contamination from lower-redshift objects and artifacts is discussed in Section~\ref{subsec:issues_photo}. An analysis of the modeling uncertainties of a $z \gtrsim 9$ sample, selected based on redshift posterior distributions contained in the catalog of this paper, is presented in  \citet{Wang2023:sys}. This is complementary to the study of a color-selected sample in the UNCOVER field, in which different photometric catalog and SED fitting methods are used \citep{Atek2023}.

\subsection{Stellar Masses and Other Ancillary Parameters}

In addition to photometric redshifts, we also release stellar mass and other ancillary parameters as listed in Table~\ref{tab:cat}. Stellar mass can often be robustly constrained by photometry; however, extra uncertainties are introduced by varying redshifts and magnifications. Lens models, especially for highly magnified regions, can be uncertain due to systematics \citep{Acebron2017,Bouwens2017,Meneghetti2017,Acebron2018,Atek2018,Furtak2021}. Therefore, the uncertainties in magnifications, which are not included in this catalog, can be significant (see \citealt{Furtak2022} for uncertainties on the lens model used in this work). We expect to fully incorporate the lensing uncertainties in the next generations of catalogs.

\subsection{Rest-frame Colors}

\begin{figure*}
\gridline{
\fig{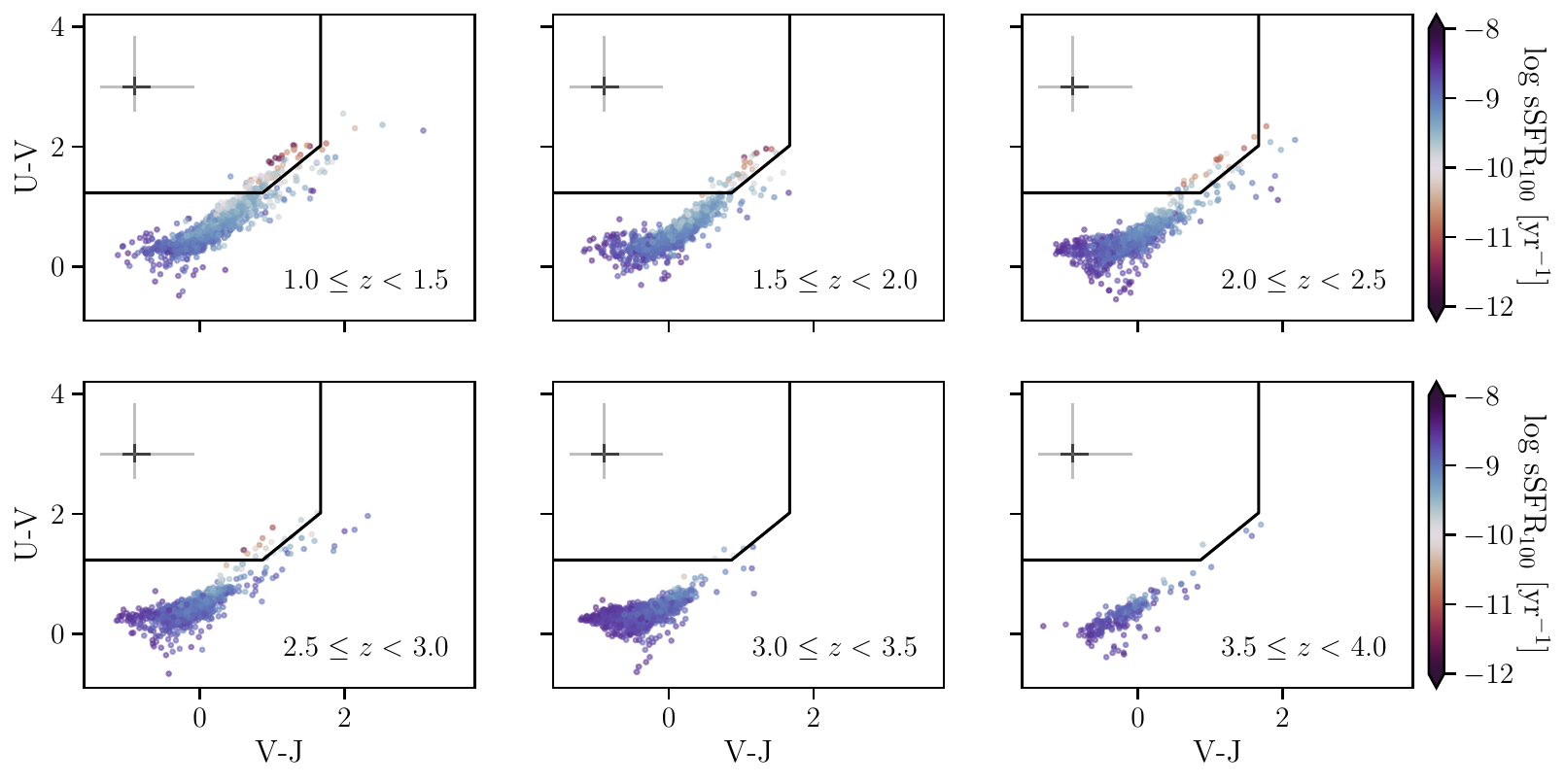}{0.99\textwidth}{}
}
\gridline{
\fig{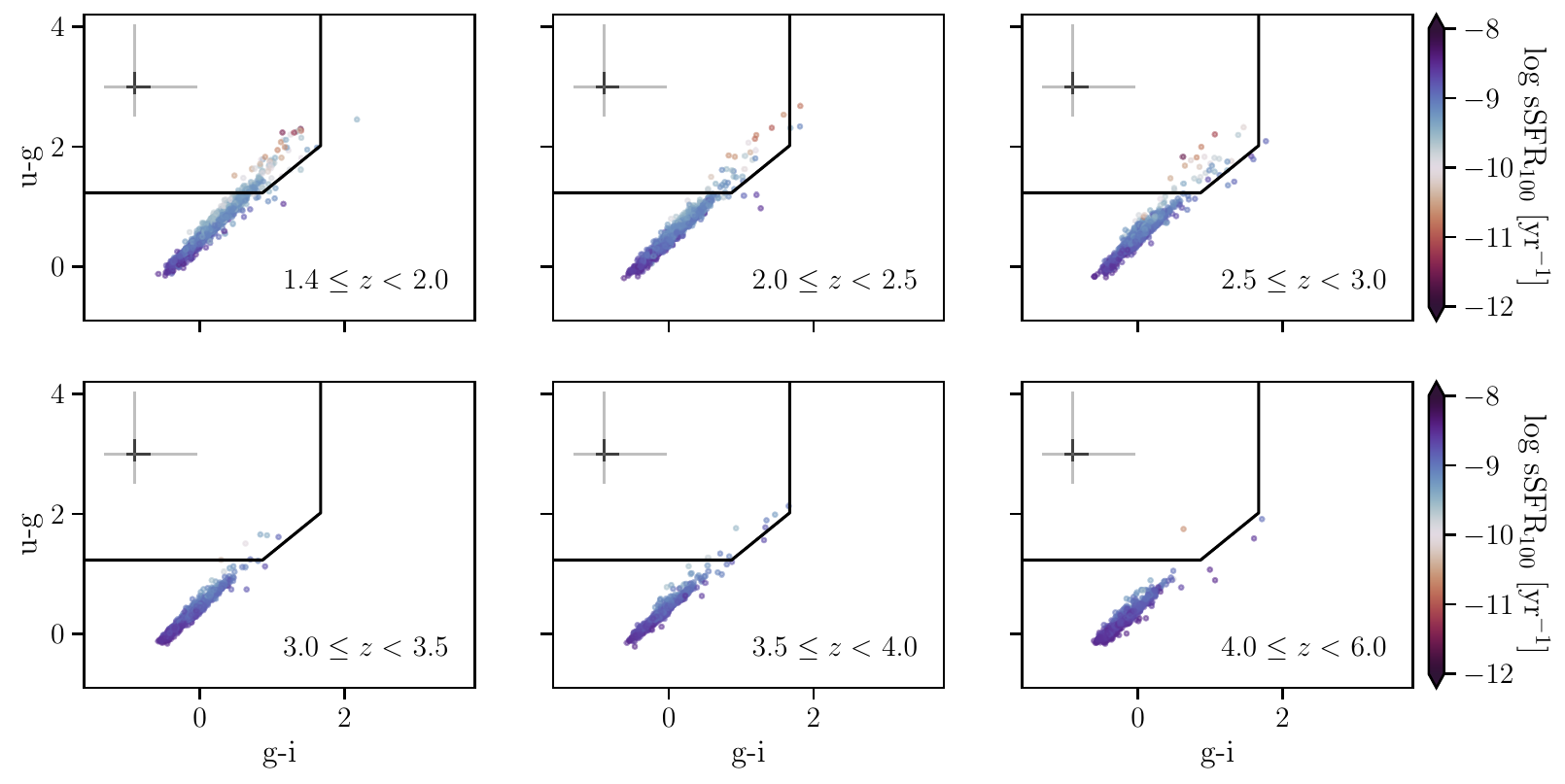}{0.99\textwidth}{}
}
\caption{Here we show the rest-frame colors from marginalizing over the \prospector-$\beta$ posteriors in redshift bins. The median and the 90\% of the uncertainty distributions are shown as black and gray error bars, respectively, in the upper left corner. The asymmetry of the error bars are driven by the non-Gaussianity of the redshift posterior distribution. The galaxies are color-coded by sSFR. The synthetic ugi colors (lower panel) are notably better correlated with sSFR than the UVJ colors (upper panel) at $1.5 \lesssim z \lesssim 3$. The best-fit versions of these plots are available in Appendix \ref{app:rf}. \label{fig:uvj}}
\end{figure*}

Considering the common practice of using rest-frame colors to categorize star-forming and quiescent galaxies (e.g., \citealt{Williams2009,Brammer2011}), we provide rest-frame fluxes in UVJ filters, and also fluxes in the recently proposed synthetic ugi filters \citep{Antwi-Danso2022}, both of which are marginalized over the full \prospector-$\beta$ posteriors. We show the UVJ and ugi color-color diagrams in redshift bins where the data has constraining power on the respective colors in Figure~\ref{fig:uvj}. Specifically, we only show the UVJ colors for galaxies in the following redshift range:
\begin{itemize}
   \item $z_{\rm phot}>1$, if optical data (HST F606W/F814W) exists
   \item $z_{\rm phot}>2.3$, otherwise
   \item $z_{\rm phot}<4$
\end{itemize}
Likewise we only show the ugi colors for galaxies in the following redshift range:
\begin{itemize}
   \item $z_{\rm phot}>1.4$, if optical data (HST F606W/F814W) exists
   \item $z_{\rm phot}>3.1$, otherwise
   \item $z_{\rm phot}<6$
\end{itemize}
These colors are, however, provided for all objects in the catalog, including cases where extrapolations are necessary.

The UVJ color-selection criteria for quiescent galaxies are
\begin{equation}
  \begin{split}
  (U-V)  & > 1.23 \; \land \; (V-J)  < 1.67 \; \land \\
     (U-V) & > (V-J) \times 0.98 + 0.38,
   \end{split}
\end{equation}
whereas the synthetic ugi color-selection criteria are
\begin{equation}
  \begin{split}
    (u - g) & > 1.5 \;  \land \; (g - i) < 1.8
    \;\land \\
    (u - g) & > (g - i) \times 0.73 + 1.08.
    \end{split}
\end{equation}
Both equations are taken from \citealt{Antwi-Danso2022}.
The synthetic ugi colors are notably better correlated with sSFR at $1.5 \lesssim z \lesssim 3$, which suggests that it may be a suitably complementary diagnostic for JWST observations. We provide further discussion on the color uncertainties and the utility of rest-frame colors calculated from the best-fit model versus posterior-averaged quantities in Appendix~\ref{app:rf}.

As a validation of the photometric modeling, we select a sample of $z>2$ quiescent candidates and cross-match to a sample of spectroscopically confirmed quiescent galaxies within Abell~2744 \citep{Marchesini2023}.
We choose a redshift-dependent definition of quiescence, calculated using a mass doubling number as
\begin{equation}
    \mathcal{D}(z) = {\rm sSFR_{100}}(z) \times t_{\rm H}(z),
\end{equation}
which is the number of times the stellar mass doubles within the age of the universe at redshift $z$, $t_{\rm H}(z)$, at the current sSFR \citep{Tacchella2022}. Focusing on the $z>2$ population, we select our quiescent sample to meet all of the following criteria based on \prospector-$\beta$ outputs:
\begin{itemize}
   \item 16th percentile of the redshift posteriors $> 2$ in the \prospector-$\beta$ fits
   \item 50th percentile of the redshift posteriors $< 9$ in the \prospector-$\beta$ fits
   \item $\mathcal{D}(z) \leq 1/20$
\end{itemize}
All the candidates passing the above criteria have no data quality issues in the cutouts or SEDs upon visual inspection.

\begin{figure*}
\gridline{
\fig{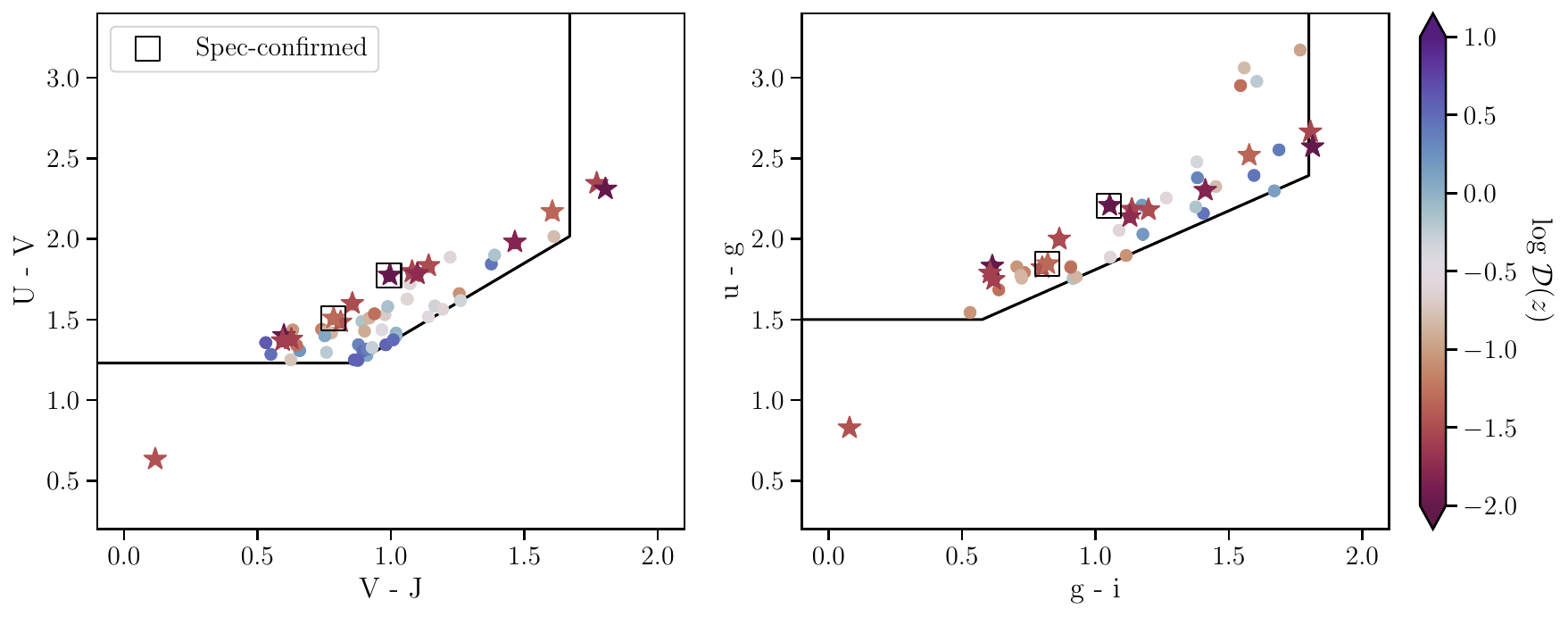}{0.99\textwidth}{}
}
\caption{All $z>2$ galaxies that satisfy the quiescent criteria of the mass doubling number $\mathcal{D}(z) \leq 1/20$ are shown as stars on the two color-color planes. Additionally, all $z>2$ galaxies that fall into the UVJ/ugi wedge but have $\mathcal{D}(z) > 1/20$ are shown as dots. The solid squares indicate the two quiescent galaxies confirmed with NIRISS spectroscopy \citep{Marchesini2023}.
\label{fig:uvj_highz_qu}}
\end{figure*}

The resulting sample is shown as stars on the UVJ and the ugi color plane in Figure~\ref{fig:uvj_highz_qu}.
We additionally mark the two spectroscopically confirmed quiescent candidates at $z > 2$ on the same figure with squares \citep{Marchesini2023}, both of which are quiescent in our photometric fits.

The implications from Figure~\ref{fig:uvj_highz_qu} are consistent with previous results in the literature: not all quiescent galaxies fall in the rest-frame color-color criteria, and not all galaxies meeting rest-frame color-color criteria are quiescent. This challenge is exacerbated at higher redshifts where all stellar populations are by definition younger and bluer \citep{Leja2019c,Carnall2023,Gould2023}. This finding also suggests that full SED-fitting can bring additional value for the selection of quiescent galaxies.

A tailored UNCOVER catalog of quiescent candidates based on different selection criteria and model assumptions will be presented in Khullar et al in prep. More dedicated discussions on quiescence can also be found therein.

\section{Known Issues in the Current DR and Upcoming Improvements\label{sec:issues}}

Thus far, this paper has presented the first-generation galaxy catalog of UNCOVER, derived from photometry that reaches intrinsic depths of $\gtrsim 31-32$ AB magnitudes after correcting for lensing magnification. We emphasize that the intention of this public release is to make the catalogs available for rapid science and is not intended to be a finalized catalog. Therefore, the user should be cautioned of the known caveats. In what follows, we discuss the known issues with the photometry and with the SPS models, as well as upcoming improvements.

\subsection{Known Challenges in the Photometric Catalog\label{subsec:issues_photo}}

This work is intended to model the photometry as given; that is, it does not attempt to solve possible issues in the first-generation photometric catalogs. Here we discuss the known issues in the context of this work; further details can be found in \citet{Weaver2023}.

Interpreting early JWST imaging data was made difficult by uncertain photometric calibration. Thanks to efforts across the community, photometric zero-points are now thought to be well understood across the detector to $<5$\% (\citealt{Boyer2022}; see also photometric calibration presented in the methods section of \citealt{Labbe2022}). 

However, it is still in the early stages of understanding and improving the known artifacts in JWST imaging \citep{Rigby2023}. These include ``claws'' and ``snowballs,'' as well as hot pixels that are particularly persistent in long-wavelength (LW) bands near the detector edges. Unidentified, these artifacts can masquerade as high-redshift galaxies, or cause spurious signals that leads to genuine high-redshift galaxies being misclassified. Such issues are difficult to identify from photometry alone, and so complicate spectral fitting. Significant effort has been expended by our team to remove these features at the reduction level; as such these issues are largely resolved in our most recent image reduction \citep{Bezanson2022} and photometric catalog \citep{Weaver2023}.

The final known issue worth discussion is that subtracting the brightest cluster galaxies leads to spurious detections of residual features. We conservatively flag objects within 3\arcsec\ of all subtracted cluster galaxy centers. However, there may remain a number of spurious sources at larger radii that are difficult to flag without simultaneously flagging robust sources. While the photometry and spectral fit of such sources may look reasonable, it is obvious when inspecting the image stamps that these sources are artifacts. While it is good practice to inspect the images of any exciting target, we especially encourage this for sources in the immediate vicinity of the three cluster cores.

\subsection{Known Unknowns in the Modeling}

SPS models require many ingredients, including an IMF, isochrones, and stellar spectra for the construction of SSPs, SFHs and stellar metallicity models for the construction of CSPs, a model for dust attenuation and emission, and nebular continuum and line emission (see \citealt{Walcher2011,Conroy2013} for recent reviews). Comprehensive assessments of SED fitting, based on data of exquisite quality and wavelength coverage at $z \lesssim 3$, have already highlighted the dependence of inferred parameters on modeling assumptions \citep{Pacifici2022}. In this work, we infer a panoramic view of the galaxy populations out to $z\sim 15$, reaching a parameter space where robust theoretical models have yet to undergo robust tests. Accordingly, we discuss a few salient points on the known unknowns in the stellar population modeling in this section.

The stellar templates are particularly uncertain at high stellar masses and at low metallicities (e.g., \citealt{Johnson2021}). The evolution of massive stars is strongly affected by the choice of input physics, including the treatment of convection, close binary evolution, rotation effects, and mass-loss, each of which has its own uncertainties \citep{Leitherer1999,Choi2016,Eldridge2017,Stanway2018,Byrne2022}. More importantly, current models assume a solar abundance pattern by necessity, but high-resolution spectra of quiescent galaxies have revealed variations in $\alpha$-element abundances \citep{Thomas2005,Choi2014,Conroy2014,Onodera2015,Kriek2016}. These trends in abundance patterns can change the fluxes by 10-40\% \citep{Vazdekis2015,Choi2019}. 
As for metallicity histories, we follow the established methodologies in the field that the same metallicity is assumed across different bins in the SFH \citep{Mitchell2013,Webb2020,Tacchella2022}. Although a possible time dependence has been claimed \citep{Bellstedt2021,Thorne2022}, we think that our approach is adequate for this work because the catalog is dominated by high-redshift, low-mass objects. These objects tends to have rising SFHs, whereas the metallicity histories are most impactful for star-forming objects with falling SFHs.
While the exact effects of the aforementioned uncertainties on SED modeling remain unclear, they can be partly illustrated by the disparate solutions found when using different stellar templates, a variety of which exist in the literature. Those templates cover different ranges of stellar evolutionary tracks and isochrones. In some cases, it has been shown that the different choices (e.g., MIST vs. PARSEC in \citealt{Whitler2022}) can lead to the inferred stellar ages differing by up to an order of magnitude. 

Nebular emission can make up a significant fraction of the total flux for stars at low metallicity and at young ages \citep{Anders2003}. This alone means that it plays an outsize importance at high redshift (e.g., \citealt{Smit2014}). Nebular emission also becomes increasingly important at high redshifts for a more technical reason: the redshifting of the spectrum causes a feature with a fixed rest-frame EW to occupy a larger fraction of the filter bandpass. It is a known challenge to model nebular emission in the early universe, due to a complex combination of non-solar abundance patterns, large uncertainties in the incident ionizing radiation, and high densities of both gas and ionizing photons \citep{Schaerer2009,Stark2013,deBarros2014,Gutkin2016,Steidel2016,Byler2017,Strom2018,Freeman2019}.
\citet{Wang2023:sys} estimate the systematic uncertainty driven by nebular physics in the inferred galaxy properties by employing a flexible nebular emission model, and find a $\sim 0.2$ dex systematic increase in stellar mass.

The initial distribution of masses for a population of stars (i.e., IMF), influences almost all the inferred properties of stellar populations---mostly notably, the total luminosity and total stellar mass. Various theories predict different shapes of the IMF (e.g., \citealt{Salpeter1955,Kroupa2001,Chabrier2003}). While it is typically assumed to be constant, there is emerging evidence that it varies across time and environment \citep{Conroy2012,2013MNRAS.433.3017L,Spiniello2014,Lyubenova2016,Lagattuta2017,vanDokkum2017}. The recently proposed temperature dependence in the IMF \citep{Steinhardt2022} and variations in a low-metallicity environment \citep{2021MNRAS.508.4175C,2022MNRAS.514.4639C} may lead to a systematic departure from the universality particularly at high redshifts. Here we assume a fixed \cite{Chabrier2003} IMF, and caution that the inferred star formation rates and stellar masses can vary by up to a factor of ten based on this assumption \citep{Wang2023:sys}. 

It is worth noting that the aforementioned systematic issues apply to all SED fitting. The inflow of JWST data breathes new life into model fitting and interpretation, as is evident from an active discussion in the literature \citep{Ferrara2022,Kannan2022,Topping2022,Adams2023,Mauerhofer2023,Mirocha2023,Reddy2023,Yung2023}. The effects of varying SPS model assumptions including burstiness in SFH, non-universal IMF, and nebular physics, using data from this catalog, are examined in detail in \citet{Wang2023:sys}.

\subsection{Reliability of Photometric Redshifts and Future Improvements}

Given the extensive discussion on inferring photometric redshifts in the literature (see \citealt{Newman2022} for a recent review, and also \citealt{Alsing2023,Leistedt2023} for general discussions on photometric redshift inference), we focus only on the unique issues faced in this work below.

Considerable uncertainties exist in the lens models, including systematic uncertainties between different models and in the highly magnified regions within a model (e.g., \citealt{Zitrin2015}).
Additionally, we note that the lensing maps do not cover our entire field of view, and thus all sources that fall outside of the lensing maps are assigned $\mu=1$. In reality the peripheral areas can be magnified by $\mu \lesssim 1.3$. A complete set of lensing maps will be released in the future. The full effect of magnification uncertainties on redshifts and other parameters will also be studied, where we add $\mu$ as a free parameter with an informative prior from the lensing maps in the SED fitting.

Most photometric redshift information comes from the position of spectral breaks, e.g., the dropout technique \citep{Steidel1996}. In principle, all high-redshift objects, with specific wavelength coverage and high S/N images taken across the multiple bands, are detectable via this technique. An obvious concern is the possible contamination from low-$z$ objects exhibiting breaks in similar locations. In particular, it can be challenging to distinguish between a Balmer break and strong emission lines \citep{Dunlop2007,Naidu2022,2023arXiv230405378A,McKinney2023,Zavala2023}. This degeneracy is mitigated, but not removed, by the adoption of the mass function prior in \prospector-$\beta$. 

Further uncertainties in the inferred photometric redshifts come from the modeling of Lyman-$\alpha$ and its interaction with the IGM. The emulator used in this work \citep{Mathews2023} is trained on a \texttt{Cloudy} grid \citep{Byler2017}, and hence does not accurately account for the radiative transfer process (e.g., damping wings). We defer a careful modeling of the Lyman-$\alpha$ to future works.

Observationally, increasing the wavelength coverage or resolution will improve the accuracy as well as the precision of the inferred redshifts. 
The Abell~2744 field will be observed with all the JWST medium bands during Cycle~2 (PI Suess, JWST-GO-4111). These observations are expected to be especially helpful in determining the uncertain emission line contribution. On the modeling side, further improvements will come from an accurate estimation of the detection efficiency as a function of mass based on the flux limits in the photometry. This will inform our model about the survey volume and downweight spurious high-redshift solutions.

Taken together, we have put our best effort forward to create the first-look catalogs. For the photometric catalogs, \citet{Weaver2023} calibrate the photometry and remove known spurious sources as cleanly as possible; whereas for the stellar population catalogs of this paper, we adopt a full Bayesian approach, incorporating empirical priors and correcting for magnification during model fitting, to ensure the maximum science return from the early observations. However, we again caution the reader to have a healthy level of skepticism in working with first-generation JWST data, especially when using sources in more exotic parameter spaces.

\section{Summary\label{sec:sum}}

In this paper, we present the first-generation galaxy catalogs spanning $0.2 \lesssim z \lesssim 15$, as part of the public release from the JWST extragalactic Treasury survey, UNCOVER. We adopt the \prospector\ Bayesian inference framework \citep{Johnson2021}, within which we use the non-parametric SFH model in \prospector-$\alpha$ \citep{Leja2019}, and three observationally motivated priors on the stellar mass functions, stellar metallicities, and SFHs in \prospector-$\beta$ \citep{Wang2023}. We constrain redshifts and galaxy properties simultaneously, meaning that the commonly non-Gaussian redshift uncertainties are propagated into inferred properties of the galaxy population. We treat lensing magnification consistently within \prospector, in contrast to the conventional approach where models are corrected for magnification post-fit.

This paper is accompanied by a catalog,
which is derived from the photometric super-catalog assembled using adaptive aperture sizes \citep{Weaver2023}. The UNCOVER SPS catalog contains the 16th, 50th, and 84th quantiles of the posterior distributions modeled with \prospector-$\beta$ for the following parameters: redshift, stellar mass, mass-weighted age, SFR, sSFR, stellar metallicity, optical depth of diffuse dust, ratio between the optical depth of birth cloud dust and diffuse dust, power law index for a \citet{Calzetti2000} attenuation curve, and ratio between the object's AGN luminosity and its bolometric luminosity.
Analytic estimates of magnifications, based on the redshift posterior medians, along with radial and tangential magnifications and shears are supplied for detailed lensing analyses \citep{Furtak2022}.
Complementary redshifts from \eazy\ are included as well.
An explanation of the catalog columns can be found in Table~\ref{tab:cat}.
We additionally provide the maximum-likelihood spectra, SFHs, and full posterior distributions.
The catalog and all related documentation are accessible via the UNCOVER survey webpage\footnote{\url{https://jwst-uncover.github.io/DR2.html\#SPSCatalogs}} with a copy deposited to Zenodo: \dataset[doi:10.5281/zenodo.8401181]{https://doi.org/10.5281/zenodo.8401181}.

Future updates to the stellar population catalog are expected in accordance with releases of extended photometric and/or spectroscopic data. The main planned enhancements in parameter inference are the incorporation of a number density prior, and lensing magnification as a free parameter with informative priors from the lens model.

We conclude by reiterating that the UNCOVER survey provides the deepest view into our universe to date by targeting the strong lens cluster Abell~2744. The stellar population catalog presented in this paper offers a rich data set to explore the processes governing galaxy formation and evolution over a redshift range now accessible by JWST.

\begin{deluxetable*}{p{0.22\textwidth} p{0.69\textwidth}}
\tablecaption{Catalog Columns\label{tab:cat}}
\tablehead{
\colhead{Column name} & \colhead{Description}
}
\startdata
\texttt{id} & unique identifier; same as in the DR2 photometric catalog \citep{Weaver2023} \\
\texttt{ra}  & RA J2000 [degrees]\\
\texttt{dec} & Dec J2000 [degrees]\\
\texttt{z\_spec} & spectroscopic redshift, where available; not including any UNCOVER MSA spec-$z$ \\
\texttt{z\_16/50/84} & redshift posterior percentiles, e.g., z16 $\to$ 16\% \\
\texttt{mstar\_16/50/84} & stellar mass [log $\msun$] \\
\texttt{mwa\_16/50/84} & mass weighted age [Gyr] \\
\texttt{sfrx\_16/50/84} & star formation rate averaged over the most recent $x~(x=10, 30, 100)$ Myr [$\msun{\rm yr^{-1}}$] \\
\texttt{ssfrx\_16/50/84} & specific star formation rate averaged over the most recent $x~(x=10, 30, 100)$ Myr [${\rm yr^{-1}}$] \\
\texttt{met\_16/50/84} & stellar metallicity [log ${\rm Z_\odot}$] \\
\texttt{dust2\_16/50/84} & optical depth of diffuse dust\\
\texttt{dust1\_fraction\_16/50/84} & ratio between the optical depth of birth cloud dust and diffuse dust \\
\texttt{dust\_index\_16/50/84} & power law index for a \citet{Calzetti2000} attenuation curve \\
\texttt{fagn\_16/50/84} & ratio between the object's AGN luminosity and its bolometric luminosity \\
\texttt{rest\_U\_16/50/84} & rest-frame $U$-band flux [AB mag] \\
\texttt{rest\_V\_16/50/84} & rest-frame $V$-band flux [AB mag] \\
\texttt{rest\_J\_16/50/84} & rest-frame $J$-band flux [AB mag] \\
\texttt{rest\_u\_16/50/84} & rest-frame synth. $u$-band flux [AB mag]\\
\texttt{rest\_g\_16/50/84} & rest-frame synth. $g$-band flux [AB mag] \\
\texttt{rest\_i\_16/50/84} & rest-frame synth. $i$-band flux [AB mag] \\
\texttt{UV\_16/50/84} & rest-frame U-V [AB mag] \\
\texttt{VJ\_16/50/84} & rest-frame V-J [AB mag] \\
\texttt{ug\_16/50/84} & rest-frame synth. u - g [AB mag] \\
\texttt{gi\_16/50/84} & rest-frame synth. g - i [AB mag] \\
\texttt{chi2} & best-fit $\chi^2$, assuming a minimum error of 5\% \\
\texttt{nbands} & number of bands used in the fit \\
\hline
\texttt{mu} & best-fit magnification based on z\_50; = 1 for foreground objects \citep{Furtak2022} \\
\texttt{mu\_16/84} & magnification uncertainty percentiles; not containing the uncertainty from $z$ posterior distributions \\
\texttt{mu\_r} & best-fit radial magnification \\
\texttt{mu\_r\_16/84} & radial magnification uncertainty percentiles  \\
\texttt{mu\_t} & best-fit tangential magnification \\
\texttt{mu\_t\_16/84} & tangential magnification uncertainty percentiles  \\
\texttt{gamma1} & best-fit shear\_1; total shear $\gamma = \sqrt{\gamma_1^2 + \gamma_1^2}$ \\
\texttt{gamma1\_16/84} & uncertainty percentiles \\
\texttt{gamma2} & best-fit shear\_2  \\
\texttt{gamma2\_16/84} & uncertainty percentiles \\
\hline
\texttt{z\_eazy} & peak of p($z$) from \eazysfhz\ \\
\texttt{z\_eazy\_16/50/84} & p($z$) percentiles \\
\hline
\texttt{id\_DR1} & ID of the source corresponding to DR1 \\
\texttt{id\_msa} & ID of the source in the MSA catalog (internal release; all within $<$~0.24\arcsec\ radius) \\
\texttt{id\_alma} & ID of ALMA source \citep{Fujimoto2023:alma} \\
\texttt{flag\_kron} & 1 for systematically underestimated photometry \\
\texttt{use\_phot} & 1 if photometry is reliable (F444W S/N $>$~3, not a star, not likely a bad pixel) \\
\texttt{use\_aper} & arcsec diameter adopted for color aperture \\
\enddata
\end{deluxetable*}

\section*{Acknowledgments}

This work is based in part on observations made with the NASA/ESA/CSA James Webb Space Telescope. The data were obtained from the Mikulski Archive for Space Telescopes at the Space Telescope Science Institute, which is operated by the Association of Universities for Research in Astronomy, Inc. (AURA), under NASA contract NAS 5-03127 for JWST. These observations are associated with JWST-GO-2561, JWST-ERS-1324, and JWST-DD-2756. Support for program JWST-GO-2561 was provided by NASA through a grant from the Space Telescope Science Institute under NASA contract NAS 5-26555. This research is also based on observations made with the NASA/ESA Hubble Space Telescope obtained from the Space Telescope Science Institute under NASA contract NAS 5–26555. These observations are associated with programs HST-GO-11689, HST-GO-13386, HST-GO/DD-13495, HST-GO-13389, HST-GO-15117, and HST-GO/DD-17231. 
The specific observations included in the JWST/UNCOVER first epoch survey
data can be accessed via
\dataset[10.17909/zn4s-0243]{https://doi.org/10.17909/zn4s-0243},
whereas the full list of HST/JWST observations used to produce our
UNCOVER catalogs is available via
\dataset[10.17909/nftp-e621]{https://doi.org/10.17909/nftp-e621}.

BW thanks the Penn State Institute for Gravitation and the Cosmos for financial support during the course of this work. RB acknowledges support from the Research Corporation for Scientific Advancement (RCSA) Cottrell Scholar Award ID No: 27587. LF and AZ acknowledge support by Grant No. 2020750 from the United States-Israel Binational Science Foundation (BSF) and Grant No.\ 2109066 from the United States National Science Foundation (NSF), and by the Ministry of Science \& Technology, Israel. PD acknowledges support from the NWO grant 016.VIDI.189.162 (``ODIN") and from the European Commission's and University of Groningen's CO-FUND Rosalind Franklin program. HA acknowledges support from CNES (Centre National d'Etudes Spatiales). RS acknowledges an STFC Ernest Rutherford Fellowship (ST/S004831/1). MS acknowledges support from the CIDEGENT/2021/059 grant, from project PID2019-109592GB-I00/AEI/10.13039/501100011033 from the Spanish Ministerio de Ciencia e Innovaci\'on - Agencia Estatal de Investigaci\'on, and from Proyecto ASFAE/2022/025 del Ministerio de Ciencia y Innovaci\'on en el marco del Plan de Recuperac\'on, Transformaci\'on y Resiliencia del Gobierno de Espa\~na. The work of CCW is supported by NOIRLab, which is managed by the AURA under a cooperative agreement with the NSF.

Computations for this research were performed on the Pennsylvania State University's Institute for Computational and Data Sciences' Roar supercomputer. This publication made use of the NASA Astrophysical Data System for bibliographic information.

\vspace{5mm}
\facilities{HST(ACS,WFC3), JWST(NIRCam, NIRSpec)}
\software{Astropy \citep{2013A&A...558A..33A,2018AJ....156..123A,2022ApJ...935..167A}, Corner \citep{2016JOSS....1...24F}, EAzY \citep{Brammer2008}, Dynesty \citep{Speagle2020}, Matplotlib \citep{2007CSE.....9...90H}, NumPy \citep{2020Natur.585..357H}, Prospector \citep{Johnson2021}, SciPy \citep{2020NatMe..17..261V}}

\appendix
\counterwithin{figure}{section}
\counterwithin{table}{section}

\section{Accuracy of Parameters Inferred with the Neural Net Emulator\label{app:is}}

\begin{figure*}
\gridline{
\fig{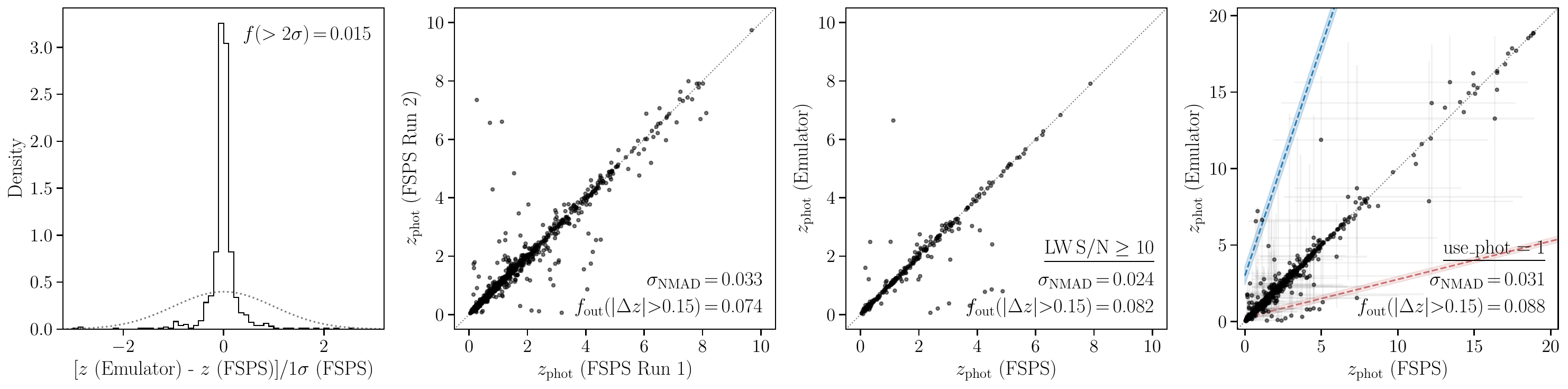}{0.99\textwidth}{}
}
\caption{Comparison between the emulator and the FSPS redshifts on a sample of 1,000 randomly drawn objects from the full UNCOVER catalog. These plots suggest that the emulator is well calibrated, and the difference is captured in the uncertainties reported in the catalog. (a) Shown here is the distribution of the difference between the emulator and FSPS redshifts normalized by the 1$\sigma$ width of the FSPS posterior distribution. A unit Gaussian is over-plotted as a dotted line. The fraction of $>2\sigma$ outliers is 0.015. (b) Medians of redshift posteriors from one run using FSPS are plotted against these from a second run under identical settings. Deviation from the diagonal reflects the sampling uncertainty arising from its probabilistic nature. (c) Medians of the redshift posteriors inferred using the emulator are plotted against the medians inferred using FSPS. Only results from a subset where S/N $\geq$ 10 in the LW detection bands are shown. (d) Same as the third panel, but the full validation set is included. Error bars are the 16th--84th quantiles. The Lyman break confusion is indicated in blue, and the shading shows $\pm$0.2 in redshift.
\label{fig:fsps_emu}}
\end{figure*}

This work is the first large-scale application of the neural net emulator for stellar population synthesis, dubbed \parrot\ \citep{Mathews2023}, which necessitates an assessment of its accuracy in inferring parameters. \citet{Mathews2023} discusses this accuracy in both flux space and in inferred parameter space but only with fixed redshifts; a free redshift significantly complexifies the parameter space by adding multiple new posterior modes, and so in this Appendix we examine the accuracy of the emulator in this new mode. We emphasize that at peak performance, the neural net emulator will exactly replicate the FSPS results, modulo any uncertainty introduced during the sampling process by the probabilistic nature of sampling.

We randomly draw 1,000 objects from the full UNCOVER sample and refit them with FSPS \citep{Conroy2010}. The accuracy of the most important and challenging parameter, redshift, is assessed in four aspects summarized in Figure~\ref{fig:fsps_emu}. First, we calculate the difference between the emulator and FSPS redshifts normalized by the 1$\sigma$ width in the FSPS posterior distribution. In general, we find the emulator performs well in this test. The great majority of the inferred photometric redshifts are within 1$\sigma$ of the full fits, with the fraction of $>2\sigma$ outliers being 0.015. Although the ideal performance is a $\delta$-function, the actual distribution of normalized residuals suggests that the emulator is well calibrated. Second, we run FSPS twice with identical settings to estimate the sampling uncertainty. The sampling method used is nested sampling \citep{Skilling2004}, which is better suited to sample multi-modal posteriors than other traditional techniques such as Markov chain Monte Carlo \citep{Goodman2010}. However, it has also been shown that nested sampling does not always accurately sample the global minimum when fitting for galaxy redshifts \citep{Wang2023b}. Our finding is consistent with this earlier study. In principle, the mode-finding problem in nested sampling can be avoided by substantial increases in the accuracy settings and thus the number of models called. This comes at a cost of increased CPU-hours, which quickly becomes prohibitively expensive for large-scale applications. Here we do not attempt to completely remove the sampling error, but instead adopt realistic \texttt{dynesty} \citep{Speagle2020} settings already more strict than those typically used in the literature, which keeps the time per fit roughly under an hour. Third, we compare the emulator and FSPS redshifts on a subsample in which the LW detection bands have S/N $\geq$~10. This allows for an examination of the emulator performance without the additional complication from data quality issues. The scatter and outlier fraction are comparable to the expectations from the sampling uncertainty, indicating that the emulator agrees well with FSPS on highly confident photometric data. Fourth, we compare the performance on the full validation set. The outlier fraction increases marginally, with the catastrophic outliers roughly follow the lines indicating the confusion between a Lyman and a Balmer break.

Given the finding above, we infer that the difference in the redshift posterior distributions found by the emulator and FSPS is mainly caused by a combination of two systematics. First, the emulated fluxes on average have 1--4\% errors compared to the exact FSPS fluxes \citep{Mathews2023}. Second, the sampler struggles to assign correct posterior mass due to the complex likelihood surface. Fitting early JWST data exacerbates this challenge; in particular in light of the newly emerged model mismatch problem suggesting photometric calibrations are yet to be complete.

With that being said, the uncertainties on the redshifts in Figure~\ref{fig:fsps_emu} accurately capture the true residuals even in the presence of emulation errors. Thus, the difference between the emulator and FSPS results are in general well described by the reported uncertainties in the catalog. An additional valuable, independent check on the fidelity of the inferred parameters can likely be obtained via an alternative inference technique, since the latter is affected by different systematics \citep{Wang2023b}, or by comparing to additional spectra. This approach will be examined in the next generations of stellar population catalogs.

\section{Photometric Residuals\label{app:residual}}

\begin{figure*}[b]
\gridline{
\fig{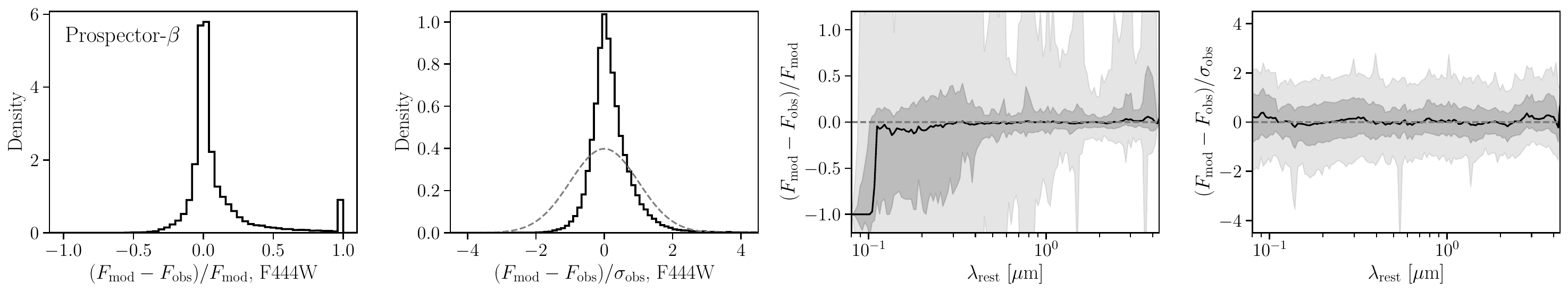}{0.99\textwidth}{}
}
\gridline{
\fig{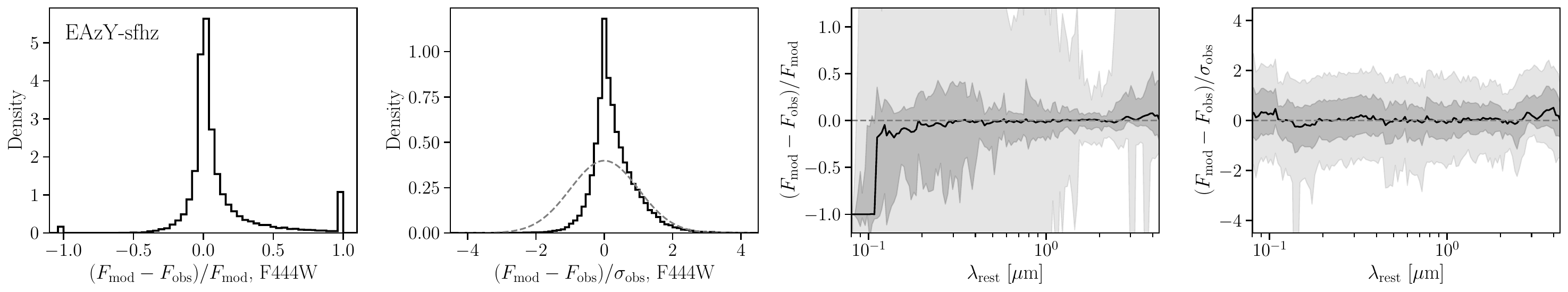}{0.99\textwidth}{}
}
\gridline{
\fig{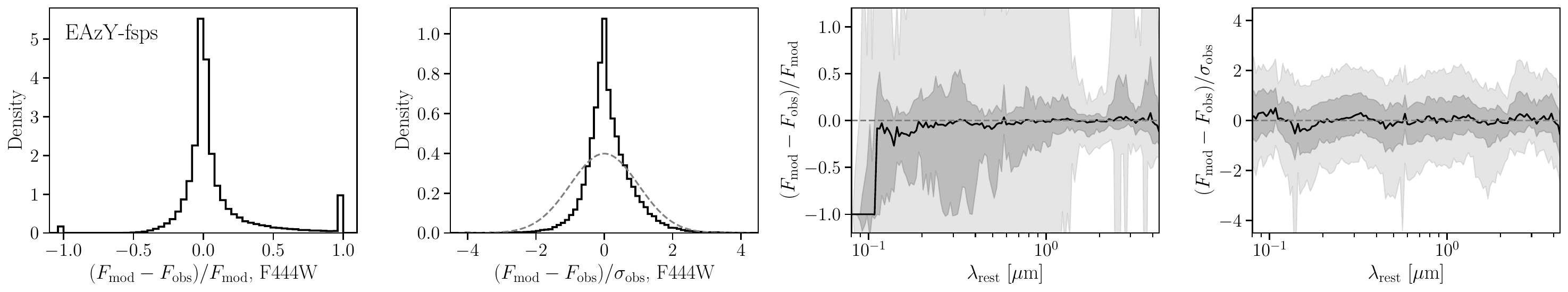}{0.99\textwidth}{}
}
\caption{The distribution of photometric residuals resulting from \prospector-$\beta$ and \eazy\ fits. We impose a minimum 5\% error floor on the photometry. The upper panel shows the following from left to right. (a) Fractional residuals in observed frame, calculated as (modeled flux-observed flux)/(modeled flux) in the F444W band; data is clipped to $\pm 1$. (b) Uncertainty-normalized residuals in observed frame, calculated as (modeled flux-observed flux)/(uncertainty in observed flux) in the F444W band; data is clipped to $\pm 3 \sigma$. The gray dashed line indicates a unit Gaussian for reference. (c) Fractional residuals in rest frame, binned logarithmically in wavelengths. (d) Error-normalized residuals in rest frame, also binned logarithmically in wavelength. Shaded regions indicate the 16--84th and the 2.5--97.5th percentiles. The middle and lower panels show the \eazysfhz\ and \eazyfsps\ residuals in the same manner. The \prospector-$\beta$ and \eazy\ fits generally agree.
However, with JWST photometric calibrations still ongoing, we caution against an over-interpolation on the performance at the current stage. 
\label{fig:residual}}
\end{figure*}

An examination of the residuals between the observed photometry and its maximum-likelihood model offers important insights into the agreement between data and the model. Figure~\ref{fig:residual} shows fractional and uncertainty-normalized residuals in the observed and the rest frames. We find general agreement between \prospector-$\beta$ and \eazy\ fits in the observed frame. Deviations from the unit Gaussian in the uncertainty-normalized residuals suggest that the photometry and the associated uncertainty require further refinement. The rest-frame residuals from \prospector-$\beta$ fits fluctuate around 0, which suggests that there is no significant bias in our model templates. The large spikes are likely due to redshift traps, since the model spectra have to be shifted by the inferred redshifts. In contrast, systematic offsets in rest-frame residuals at short wavelengths are observed in \eazysfhz\ and \eazyfsps\ fits, which may indicate template mismatches. However, given the on-going efforts in the photometric calibration, we caution against an over-interpretation of the \eazy\ performance at the current stage.

\section{Rest-frame Colors\label{app:rf}}

\begin{figure*}
\gridline{
\fig{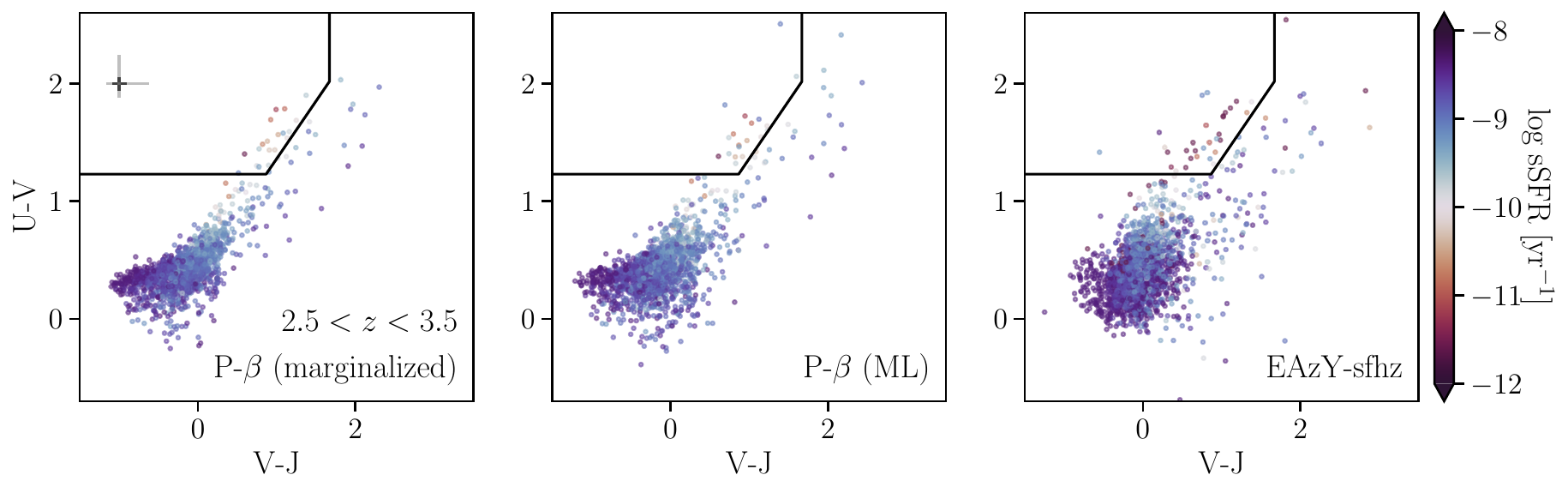}{0.99\textwidth}{}
}
\caption{Here we compare the rest-frame colors of galaxies at $2.5<z<3.5$ inferred by three different methods: marginalizing over the full \prospector-$\beta$ posteriors, colors inferred from the maximum-likelihood (ML) spectra from \prospector-$\beta$, and from the best-fit templates of \eazysfhz. The medians of the uncertainty distributions are indicated by black error bars, while the 90\% quantiles of the uncertainty distributions are indicated by gray error bars. The asymmetry of the error bars is driven by the non-Gaussianity of the redshift posterior distribution. Only objects of which S/N $\geq$ 10 in the LW detection bands are included for the ease of interpreting the difference. The marginalized colors have a smaller scatter because marginalization tamps down the photometric uncertainties, but their uncertainties are reflective of the larger scatters found by the other two methods.
\label{fig:uvj_cali}}
\end{figure*}

The rest-frame colors released in the catalogs are marginalized over the full \prospector-$\beta$ posteriors. Here we additionally show the colors estimated from maximum-likelihood spectra in Figure~\ref{fig:uvj_cali}, which have larger scatter as expected. We also include the colors from \eazysfhz. \eazy\ measures the colors directly from the best-fit template, which is simply a linear combination of templates that minimizes the $\chi^2$ statistics. Less information is encoded in the process determining the colors this way; it is thus understandable that the \eazy\ colors exhibit a larger scatter.

\bibliography{uncover_sps_wang.bib}

\begin{thebibliography}{}
\expandafter\ifx\csname natexlab\endcsname\relax\def\natexlab#1{#1}\fi
\providecommand{\url}[1]{\href{#1}{#1}}
\providecommand{\dodoi}[1]{doi:~\href{http://doi.org/#1}{\nolinkurl{#1}}}
\providecommand{\doeprint}[1]{\href{http://ascl.net/#1}{\nolinkurl{http://ascl.net/#1}}}
\providecommand{\doarXiv}[1]{\href{https://arxiv.org/abs/#1}{\nolinkurl{https://arxiv.org/abs/#1}}}

\bibitem[{{Acebron} {et~al.}(2017){Acebron}, {Jullo}, {Limousin}, {Tilquin},
  {Giocoli}, {Jauzac}, {Mahler}, \& {Richard}}]{Acebron2017}
{Acebron}, A., {Jullo}, E., {Limousin}, M., {et~al.} 2017, \mnras, 470, 1809,
  \dodoi{10.1093/mnras/stx1330}

\bibitem[{{Acebron} {et~al.}(2018){Acebron}, {Cibirka}, {Zitrin}, {Coe},
  {Agulli}, {Sharon}, {Brada{\v{c}}}, {Frye}, {Livermore}, {Mahler}, {Salmon},
  {Umetsu}, {Bradley}, {Andrade-Santos}, {Avila}, {Carrasco}, {Cerny},
  {Czakon}, {Dawson}, {Hoag}, {Huang}, {Johnson}, {Jones}, {Kikuchihara},
  {Lam}, {Lovisari}, {Mainali}, {Oesch}, {Ogaz}, {Ouchi}, {Past},
  {Paterno-Mahler}, {Peterson}, {Ryan}, {Sendra-Server}, {Stark}, {Strait},
  {Toft}, {Trenti}, \& {Vulcani}}]{Acebron2018}
{Acebron}, A., {Cibirka}, N., {Zitrin}, A., {et~al.} 2018, \apj, 858, 42,
  \dodoi{10.3847/1538-4357/aabe29}

\bibitem[{{Adams} {et~al.}(2023){Adams}, {Conselice}, {Ferreira}, {Austin},
  {Trussler}, {Juod{\v{z}}balis}, {Wilkins}, {Caruana}, {Dayal}, {Verma}, \&
  {Vijayan}}]{Adams2023}
{Adams}, N.~J., {Conselice}, C.~J., {Ferreira}, L., {et~al.} 2023, \mnras, 518,
  4755, \dodoi{10.1093/mnras/stac3347}

\bibitem[{{Alsing} {et~al.}(2023){Alsing}, {Peiris}, {Mortlock}, {Leja}, \&
  {Leistedt}}]{Alsing2023}
{Alsing}, J., {Peiris}, H., {Mortlock}, D., {Leja}, J., \& {Leistedt}, B. 2023,
  \apjs, 264, 29, \dodoi{10.3847/1538-4365/ac9583}

\bibitem[{{Alsing} {et~al.}(2020){Alsing}, {Peiris}, {Leja}, {Hahn}, {Tojeiro},
  {Mortlock}, {Leistedt}, {Johnson}, \& {Conroy}}]{Alsing2020}
{Alsing}, J., {Peiris}, H., {Leja}, J., {et~al.} 2020, \apjs, 249, 5,
  \dodoi{10.3847/1538-4365/ab917f}

\bibitem[{{Anders} \& {Fritze-v. Alvensleben}(2003)}]{Anders2003}
{Anders}, P., \& {Fritze-v. Alvensleben}, U. 2003, \aap, 401, 1063,
  \dodoi{10.1051/0004-6361:20030151}

\bibitem[{{Antwi-Danso} {et~al.}(2023){Antwi-Danso}, {Papovich}, {Leja},
  {Marchesini}, {Marsan}, {Martis}, {Labb{\'e}}, {Muzzin}, {Glazebrook},
  {Straatman}, \& {Tran}}]{Antwi-Danso2022}
{Antwi-Danso}, J., {Papovich}, C., {Leja}, J., {et~al.} 2023, \apj, 943, 166,
  \dodoi{10.3847/1538-4357/aca294}

\bibitem[{{Arrabal Haro} {et~al.}(2023){Arrabal Haro}, {Dickinson},
  {Finkelstein}, {Fujimoto}, {Fern{\'a}ndez}, {Kartaltepe}, {Jung}, {Cole},
  {Burgarella}, {Chworowsky}, {Hutchison}, {Morales}, {Papovich}, {Simons},
  {Amor{\'\i}n}, {Backhaus}, {Bagley}, {Bisigello}, {Calabr{\`o}},
  {Castellano}, {Cleri}, {Dav{\'e}}, {Dekel}, {Ferguson}, {Fontana}, {Gawiser},
  {Giavalisco}, {Harish}, {Hathi}, {Hirschmann}, {Holwerda}, {Huertas-Company},
  {Koekemoer}, {Larson}, {Lucas}, {Mobasher}, {P{\'e}rez-Gonz{\'a}lez},
  {Pirzkal}, {Rose}, {Santini}, {Trump}, {de la Vega}, {Wang}, {Weiner},
  {Wilkins}, {Yang}, {Yung}, \& {Zavala}}]{2023arXiv230405378A}
{Arrabal Haro}, P., {Dickinson}, M., {Finkelstein}, S.~L., {et~al.} 2023,
  \apjl, 951, L22, \dodoi{10.3847/2041-8213/acdd54}

\bibitem[{{Astropy Collaboration} {et~al.}(2013){Astropy Collaboration},
  {Robitaille}, {Tollerud}, {Greenfield}, {Droettboom}, {Bray}, {Aldcroft},
  {Davis}, {Ginsburg}, {Price-Whelan}, {Kerzendorf}, {Conley}, {Crighton},
  {Barbary}, {Muna}, {Ferguson}, {Grollier}, {Parikh}, {Nair}, {Unther},
  {Deil}, {Woillez}, {Conseil}, {Kramer}, {Turner}, {Singer}, {Fox}, {Weaver},
  {Zabalza}, {Edwards}, {Azalee Bostroem}, {Burke}, {Casey}, {Crawford},
  {Dencheva}, {Ely}, {Jenness}, {Labrie}, {Lim}, {Pierfederici}, {Pontzen},
  {Ptak}, {Refsdal}, {Servillat}, \& {Streicher}}]{2013A&A...558A..33A}
{Astropy Collaboration}, {Robitaille}, T.~P., {Tollerud}, E.~J., {et~al.} 2013,
  \aap, 558, A33, \dodoi{10.1051/0004-6361/201322068}

\bibitem[{{Astropy Collaboration} {et~al.}(2018){Astropy Collaboration},
  {Price-Whelan}, {Sip{\H{o}}cz}, {G{\"u}nther}, {Lim}, {Crawford}, {Conseil},
  {Shupe}, {Craig}, {Dencheva}, {Ginsburg}, {VanderPlas}, {Bradley},
  {P{\'e}rez-Su{\'a}rez}, {de Val-Borro}, {Aldcroft}, {Cruz}, {Robitaille},
  {Tollerud}, {Ardelean}, {Babej}, {Bach}, {Bachetti}, {Bakanov}, {Bamford},
  {Barentsen}, {Barmby}, {Baumbach}, {Berry}, {Biscani}, {Boquien}, {Bostroem},
  {Bouma}, {Brammer}, {Bray}, {Breytenbach}, {Buddelmeijer}, {Burke},
  {Calderone}, {Cano Rodr{\'\i}guez}, {Cara}, {Cardoso}, {Cheedella}, {Copin},
  {Corrales}, {Crichton}, {D'Avella}, {Deil}, {Depagne}, {Dietrich}, {Donath},
  {Droettboom}, {Earl}, {Erben}, {Fabbro}, {Ferreira}, {Finethy}, {Fox},
  {Garrison}, {Gibbons}, {Goldstein}, {Gommers}, {Greco}, {Greenfield},
  {Groener}, {Grollier}, {Hagen}, {Hirst}, {Homeier}, {Horton}, {Hosseinzadeh},
  {Hu}, {Hunkeler}, {Ivezi{\'c}}, {Jain}, {Jenness}, {Kanarek}, {Kendrew},
  {Kern}, {Kerzendorf}, {Khvalko}, {King}, {Kirkby}, {Kulkarni}, {Kumar},
  {Lee}, {Lenz}, {Littlefair}, {Ma}, {Macleod}, {Mastropietro}, {McCully},
  {Montagnac}, {Morris}, {Mueller}, {Mumford}, {Muna}, {Murphy}, {Nelson},
  {Nguyen}, {Ninan}, {N{\"o}the}, {Ogaz}, {Oh}, {Parejko}, {Parley}, {Pascual},
  {Patil}, {Patil}, {Plunkett}, {Prochaska}, {Rastogi}, {Reddy Janga},
  {Sabater}, {Sakurikar}, {Seifert}, {Sherbert}, {Sherwood-Taylor}, {Shih},
  {Sick}, {Silbiger}, {Singanamalla}, {Singer}, {Sladen}, {Sooley},
  {Sornarajah}, {Streicher}, {Teuben}, {Thomas}, {Tremblay}, {Turner},
  {Terr{\'o}n}, {van Kerkwijk}, {de la Vega}, {Watkins}, {Weaver}, {Whitmore},
  {Woillez}, {Zabalza}, \& {Astropy Contributors}}]{2018AJ....156..123A}
{Astropy Collaboration}, {Price-Whelan}, A.~M., {Sip{\H{o}}cz}, B.~M., {et~al.}
  2018, \aj, 156, 123, \dodoi{10.3847/1538-3881/aabc4f}

\bibitem[{{Astropy Collaboration} {et~al.}(2022){Astropy Collaboration},
  {Price-Whelan}, {Lim}, {Earl}, {Starkman}, {Bradley}, {Shupe}, {Patil},
  {Corrales}, {Brasseur}, {N{\"o}the}, {Donath}, {Tollerud}, {Morris},
  {Ginsburg}, {Vaher}, {Weaver}, {Tocknell}, {Jamieson}, {van Kerkwijk},
  {Robitaille}, {Merry}, {Bachetti}, {G{\"u}nther}, {Aldcroft},
  {Alvarado-Montes}, {Archibald}, {B{\'o}di}, {Bapat}, {Barentsen},
  {Baz{\'a}n}, {Biswas}, {Boquien}, {Burke}, {Cara}, {Cara}, {Conroy},
  {Conseil}, {Craig}, {Cross}, {Cruz}, {D'Eugenio}, {Dencheva}, {Devillepoix},
  {Dietrich}, {Eigenbrot}, {Erben}, {Ferreira}, {Foreman-Mackey}, {Fox},
  {Freij}, {Garg}, {Geda}, {Glattly}, {Gondhalekar}, {Gordon}, {Grant},
  {Greenfield}, {Groener}, {Guest}, {Gurovich}, {Handberg}, {Hart},
  {Hatfield-Dodds}, {Homeier}, {Hosseinzadeh}, {Jenness}, {Jones}, {Joseph},
  {Kalmbach}, {Karamehmetoglu}, {Ka{\l}uszy{\'n}ski}, {Kelley}, {Kern},
  {Kerzendorf}, {Koch}, {Kulumani}, {Lee}, {Ly}, {Ma}, {MacBride}, {Maljaars},
  {Muna}, {Murphy}, {Norman}, {O'Steen}, {Oman}, {Pacifici}, {Pascual},
  {Pascual-Granado}, {Patil}, {Perren}, {Pickering}, {Rastogi}, {Roulston},
  {Ryan}, {Rykoff}, {Sabater}, {Sakurikar}, {Salgado}, {Sanghi}, {Saunders},
  {Savchenko}, {Schwardt}, {Seifert-Eckert}, {Shih}, {Jain}, {Shukla}, {Sick},
  {Simpson}, {Singanamalla}, {Singer}, {Singhal}, {Sinha}, {Sip{\H{o}}cz},
  {Spitler}, {Stansby}, {Streicher}, {{\v{S}}umak}, {Swinbank}, {Taranu},
  {Tewary}, {Tremblay}, {de Val-Borro}, {Van Kooten}, {Vasovi{\'c}}, {Verma},
  {de Miranda Cardoso}, {Williams}, {Wilson}, {Winkel}, {Wood-Vasey}, {Xue},
  {Yoachim}, {Zhang}, {Zonca}, \& {Astropy Project
  Contributors}}]{2022ApJ...935..167A}
{Astropy Collaboration}, {Price-Whelan}, A.~M., {Lim}, P.~L., {et~al.} 2022,
  \apj, 935, 167, \dodoi{10.3847/1538-4357/ac7c74}

\bibitem[{{Atek} {et~al.}(2018){Atek}, {Richard}, {Kneib}, \&
  {Schaerer}}]{Atek2018}
{Atek}, H., {Richard}, J., {Kneib}, J.-P., \& {Schaerer}, D. 2018, \mnras, 479,
  5184, \dodoi{10.1093/mnras/sty1820}

\bibitem[{{Atek} {et~al.}(2023){Atek}, {Chemerynska}, {Wang}, {Furtak},
  {Weibel}, {Oesch}, {Weaver}, {Labb{\'e}}, {Bezanson}, {van Dokkum}, {Zitrin},
  {Dayal}, {Williams}, {Nannayakkara}, {Price}, {Brammer}, {Goulding}, {Leja},
  {Marchesini}, {Nelson}, {Pan}, \& {Whitaker}}]{Atek2023}
{Atek}, H., {Chemerynska}, I., {Wang}, B., {et~al.} 2023, \mnras, 524, 5486,
  \dodoi{10.1093/mnras/stad1998}

\bibitem[{{Behroozi} {et~al.}(2019){Behroozi}, {Wechsler}, {Hearin}, \&
  {Conroy}}]{Behroozi2019}
{Behroozi}, P., {Wechsler}, R.~H., {Hearin}, A.~P., \& {Conroy}, C. 2019,
  \mnras, 488, 3143, \dodoi{10.1093/mnras/stz1182}

\bibitem[{{Bellstedt} {et~al.}(2021){Bellstedt}, {Robotham}, {Driver},
  {Thorne}, {Davies}, {Holwerda}, {Hopkins}, {Lara-Lopez},
  {L{\'o}pez-S{\'a}nchez}, \& {Phillipps}}]{Bellstedt2021}
{Bellstedt}, S., {Robotham}, A. S.~G., {Driver}, S.~P., {et~al.} 2021, \mnras,
  503, 3309, \dodoi{10.1093/mnras/stab550}

\bibitem[{{Bezanson} {et~al.}(2022){Bezanson}, {Labbe}, {Whitaker}, {Leja},
  {Price}, {Franx}, {Brammer}, {Marchesini}, {Zitrin}, {Wang}, {Weaver},
  {Furtak}, {Atek}, {Coe}, {Cutler}, {Dayal}, {van Dokkum}, {Feldmann},
  {Forster Schreiber}, {Fujimoto}, {Geha}, {Glazebrook}, {de Graaff}, {Greene},
  {Juneau}, {Kassin}, {Kriek}, {Khullar}, {Maseda}, {Mowla}, {Muzzin},
  {Nanayakkara}, {Nelson}, {Oesch}, {Pacifici}, {Pan}, {Papovich}, {Setton},
  {Shapley}, {Smit}, {Stefanon}, {Taylor}, \& {Williams}}]{Bezanson2022}
{Bezanson}, R., {Labbe}, I., {Whitaker}, K.~E., {et~al.} 2022, arXiv e-prints,
  arXiv:2212.04026.
\newblock \doarXiv{2212.04026}

\bibitem[{{B{\"o}ker} {et~al.}(2023){B{\"o}ker}, {Beck}, {Birkmann},
  {Giardino}, {Keyes}, {Kumari}, {Muzerolle}, {Rawle}, {Zeidler}, {Abul-Huda},
  {Alves de Oliveira}, {Arribas}, {Bechtold}, {Bhatawdekar}, {Bonaventura},
  {Bunker}, {Cameron}, {Carniani}, {Charlot}, {Curti}, {Espinoza}, {Ferruit},
  {Franx}, {Jakobsen}, {Karakla}, {L{\'o}pez-Caniego}, {L{\"u}tzgendorf},
  {Maiolino}, {Manjavacas}, {Marston}, {Moseley}, {Ogle}, {Perna},
  {Pe{\~n}a-Guerrero}, {Pirzkal}, {Plesha}, {Proffitt}, {Rauscher}, {Rix},
  {Rodr{\'\i}guez del Pino}, {Rustamkulov}, {Sabbi}, {Sing}, {Sirianni}, {te
  Plate}, {{\'U}beda}, {Wahlgren}, {Wislowski}, {Wu}, \& {Willott}}]{Boker2023}
{B{\"o}ker}, T., {Beck}, T.~L., {Birkmann}, S.~M., {et~al.} 2023, \pasp, 135,
  038001, \dodoi{10.1088/1538-3873/acb846}

\bibitem[{{Bouwens} {et~al.}(2017){Bouwens}, {Oesch}, {Illingworth}, {Ellis},
  \& {Stefanon}}]{Bouwens2017}
{Bouwens}, R.~J., {Oesch}, P.~A., {Illingworth}, G.~D., {Ellis}, R.~S., \&
  {Stefanon}, M. 2017, \apj, 843, 129, \dodoi{10.3847/1538-4357/aa70a4}

\bibitem[{{Boyer} {et~al.}(2022){Boyer}, {Anderson}, {Gennaro}, {Geha},
  {Wingfield McQuinn}, {Tollerud}, {Correnti}, {Brenner Newman}, {Cohen},
  {Kallivayalil}, {Beaton}, {Cole}, {Dolphin}, {Kalirai}, {Sandstrom},
  {Savino}, {Skillman}, {Weisz}, \& {Williams}}]{Boyer2022}
{Boyer}, M.~L., {Anderson}, J., {Gennaro}, M., {et~al.} 2022, Research Notes of
  the American Astronomical Society, 6, 191, \dodoi{10.3847/2515-5172/ac923a}

\bibitem[{{Brammer} {et~al.}(2008){Brammer}, {van Dokkum}, \&
  {Coppi}}]{Brammer2008}
{Brammer}, G.~B., {van Dokkum}, P.~G., \& {Coppi}, P. 2008, \apj, 686, 1503,
  \dodoi{10.1086/591786}

\bibitem[{{Brammer} {et~al.}(2011){Brammer}, {Whitaker}, {van Dokkum},
  {Marchesini}, {Franx}, {Kriek}, {Labb{\'e}}, {Lee}, {Muzzin}, {Quadri},
  {Rudnick}, \& {Williams}}]{Brammer2011}
{Brammer}, G.~B., {Whitaker}, K.~E., {van Dokkum}, P.~G., {et~al.} 2011, \apj,
  739, 24, \dodoi{10.1088/0004-637X/739/1/24}

\bibitem[{{Brammer} {et~al.}(2012){Brammer}, {S{\'a}nchez-Janssen},
  {Labb{\'e}}, {da Cunha}, {Erb}, {Franx}, {Fumagalli}, {Lundgren},
  {Marchesini}, {Momcheva}, {Nelson}, {Patel}, {Quadri}, {Rix}, {Skelton},
  {Schmidt}, {van der Wel}, {van Dokkum}, {Wake}, \&
  {Whitaker}}]{2012ApJ...758L..17B}
{Brammer}, G.~B., {S{\'a}nchez-Janssen}, R., {Labb{\'e}}, I., {et~al.} 2012,
  \apjl, 758, L17, \dodoi{10.1088/2041-8205/758/1/L17}

\bibitem[{{Byler} {et~al.}(2017){Byler}, {Dalcanton}, {Conroy}, \&
  {Johnson}}]{Byler2017}
{Byler}, N., {Dalcanton}, J.~J., {Conroy}, C., \& {Johnson}, B.~D. 2017, \apj,
  840, 44, \dodoi{10.3847/1538-4357/aa6c66}

\bibitem[{{Byrne} {et~al.}(2022){Byrne}, {Stanway}, {Eldridge}, {McSwiney}, \&
  {Townsend}}]{Byrne2022}
{Byrne}, C.~M., {Stanway}, E.~R., {Eldridge}, J.~J., {McSwiney}, L., \&
  {Townsend}, O.~T. 2022, \mnras, 512, 5329, \dodoi{10.1093/mnras/stac807}

\bibitem[{{Calzetti} {et~al.}(2000){Calzetti}, {Armus}, {Bohlin}, {Kinney},
  {Koornneef}, \& {Storchi-Bergmann}}]{Calzetti2000}
{Calzetti}, D., {Armus}, L., {Bohlin}, R.~C., {et~al.} 2000, \apj, 533, 682,
  \dodoi{10.1086/308692}

\bibitem[{{Carnall} {et~al.}(2023){Carnall}, {Begley}, {McLeod}, {Hamadouche},
  {Donnan}, {McLure}, {Dunlop}, {Milvang-Jensen}, {Bondestam}, {Cullen},
  {Jewell}, \& {Pollock}}]{Carnall2023}
{Carnall}, A.~C., {Begley}, R., {McLeod}, D.~J., {et~al.} 2023, \mnras, 518,
  L45, \dodoi{10.1093/mnrasl/slac136}

\bibitem[{{Chabrier}(2003)}]{Chabrier2003}
{Chabrier}, G. 2003, \pasp, 115, 763, \dodoi{10.1086/376392}

\bibitem[{{Charlot} \& {Fall}(2000)}]{Charlot2000}
{Charlot}, S., \& {Fall}, S.~M. 2000, \apj, 539, 718, \dodoi{10.1086/309250}

\bibitem[{{Choi} {et~al.}(2019){Choi}, {Conroy}, \& {Johnson}}]{Choi2019}
{Choi}, J., {Conroy}, C., \& {Johnson}, B.~D. 2019, \apj, 872, 136,
  \dodoi{10.3847/1538-4357/aaff67}

\bibitem[{{Choi} {et~al.}(2014){Choi}, {Conroy}, {Moustakas}, {Graves},
  {Holden}, {Brodwin}, {Brown}, \& {van Dokkum}}]{Choi2014}
{Choi}, J., {Conroy}, C., {Moustakas}, J., {et~al.} 2014, \apj, 792, 95,
  \dodoi{10.1088/0004-637X/792/2/95}

\bibitem[{{Choi} {et~al.}(2016){Choi}, {Dotter}, {Conroy}, {Cantiello},
  {Paxton}, \& {Johnson}}]{Choi2016}
{Choi}, J., {Dotter}, A., {Conroy}, C., {et~al.} 2016, \apj, 823, 102,
  \dodoi{10.3847/0004-637X/823/2/102}

\bibitem[{{Chon} {et~al.}(2021){Chon}, {Omukai}, \&
  {Schneider}}]{2021MNRAS.508.4175C}
{Chon}, S., {Omukai}, K., \& {Schneider}, R. 2021, \mnras, 508, 4175,
  \dodoi{10.1093/mnras/stab2497}

\bibitem[{{Chon} {et~al.}(2022){Chon}, {Ono}, {Omukai}, \&
  {Schneider}}]{2022MNRAS.514.4639C}
{Chon}, S., {Ono}, H., {Omukai}, K., \& {Schneider}, R. 2022, \mnras, 514,
  4639, \dodoi{10.1093/mnras/stac1549}

\bibitem[{{Conroy}(2013)}]{Conroy2013}
{Conroy}, C. 2013, \araa, 51, 393, \dodoi{10.1146/annurev-astro-082812-141017}

\bibitem[{{Conroy} {et~al.}(2014){Conroy}, {Graves}, \& {van
  Dokkum}}]{Conroy2014}
{Conroy}, C., {Graves}, G.~J., \& {van Dokkum}, P.~G. 2014, \apj, 780, 33,
  \dodoi{10.1088/0004-637X/780/1/33}

\bibitem[{{Conroy} \& {Gunn}(2010)}]{Conroy2010}
{Conroy}, C., \& {Gunn}, J.~E. 2010, \apj, 712, 833,
  \dodoi{10.1088/0004-637X/712/2/833}

\bibitem[{{Conroy} \& {van Dokkum}(2012)}]{Conroy2012}
{Conroy}, C., \& {van Dokkum}, P.~G. 2012, \apj, 760, 71,
  \dodoi{10.1088/0004-637X/760/1/71}

\bibitem[{{Cowie} {et~al.}(1996){Cowie}, {Songaila}, {Hu}, \&
  {Cohen}}]{Cowie1996}
{Cowie}, L.~L., {Songaila}, A., {Hu}, E.~M., \& {Cohen}, J.~G. 1996, \aj, 112,
  839, \dodoi{10.1086/118058}

\bibitem[{{de Barros} {et~al.}(2014){de Barros}, {Schaerer}, \&
  {Stark}}]{deBarros2014}
{de Barros}, S., {Schaerer}, D., \& {Stark}, D.~P. 2014, \aap, 563, A81,
  \dodoi{10.1051/0004-6361/201220026}

\bibitem[{{Dotter}(2016)}]{Dotter2016}
{Dotter}, A. 2016, \apjs, 222, 8, \dodoi{10.3847/0067-0049/222/1/8}

\bibitem[{{Draine} \& {Li}(2007)}]{Draine2007}
{Draine}, B.~T., \& {Li}, A. 2007, \apj, 657, 810, \dodoi{10.1086/511055}

\bibitem[{{Dunlop} {et~al.}(2007){Dunlop}, {Cirasuolo}, \&
  {McLure}}]{Dunlop2007}
{Dunlop}, J.~S., {Cirasuolo}, M., \& {McLure}, R.~J. 2007, \mnras, 376, 1054,
  \dodoi{10.1111/j.1365-2966.2007.11453.x}

\bibitem[{{Eldridge} {et~al.}(2017){Eldridge}, {Stanway}, {Xiao}, {McClelland},
  {Taylor}, {Ng}, {Greis}, \& {Bray}}]{Eldridge2017}
{Eldridge}, J.~J., {Stanway}, E.~R., {Xiao}, L., {et~al.} 2017, \pasa, 34,
  e058, \dodoi{10.1017/pasa.2017.51}

\bibitem[{{Feldmann} {et~al.}(2023){Feldmann}, {Quataert},
  {Faucher-Gigu{\`e}re}, {Hopkins}, {{\c{C}}atmabacak}, {Kere{\v{s}}},
  {Bassini}, {Bernardini}, {Bullock}, {Cenci}, {Gensior}, {Liang}, {Moreno}, \&
  {Wetzel}}]{Feldmann2023}
{Feldmann}, R., {Quataert}, E., {Faucher-Gigu{\`e}re}, C.-A., {et~al.} 2023,
  \mnras, 522, 3831, \dodoi{10.1093/mnras/stad1205}

\bibitem[{{Ferrara} {et~al.}(2023){Ferrara}, {Pallottini}, \&
  {Dayal}}]{Ferrara2022}
{Ferrara}, A., {Pallottini}, A., \& {Dayal}, P. 2023, \mnras, 522, 3986,
  \dodoi{10.1093/mnras/stad1095}

\bibitem[{{Finkelstein} {et~al.}(2023){Finkelstein}, {Bagley}, {Ferguson},
  {Wilkins}, {Kartaltepe}, {Papovich}, {Yung}, {Arrabal Haro}, {Behroozi},
  {Dickinson}, {Kocevski}, {Koekemoer}, {Larson}, {Le Bail}, {Morales},
  {P{\'e}rez-Gonz{\'a}lez}, {Burgarella}, {Dav{\'e}}, {Hirschmann},
  {Somerville}, {Wuyts}, {Bromm}, {Casey}, {Fontana}, {Fujimoto}, {Gardner},
  {Giavalisco}, {Grazian}, {Grogin}, {Hathi}, {Hutchison}, {Jha}, {Jogee},
  {Kewley}, {Kirkpatrick}, {Long}, {Lotz}, {Pentericci}, {Pierel}, {Pirzkal},
  {Ravindranath}, {Ryan}, {Trump}, {Yang}, {Bhatawdekar}, {Bisigello}, {Buat},
  {Calabr{\`o}}, {Castellano}, {Cleri}, {Cooper}, {Croton}, {Daddi}, {Dekel},
  {Elbaz}, {Franco}, {Gawiser}, {Holwerda}, {Huertas-Company}, {Jaskot},
  {Leung}, {Lucas}, {Mobasher}, {Pandya}, {Tacchella}, {Weiner}, \&
  {Zavala}}]{2022arXiv221105792F}
{Finkelstein}, S.~L., {Bagley}, M.~B., {Ferguson}, H.~C., {et~al.} 2023, \apjl,
  946, L13, \dodoi{10.3847/2041-8213/acade4}

\bibitem[{{Foreman-Mackey}(2016)}]{2016JOSS....1...24F}
{Foreman-Mackey}, D. 2016, JOSS, 1, 24, \dodoi{10.21105/joss.00024}

\bibitem[{{Freeman} {et~al.}(2019){Freeman}, {Siana}, {Kriek}, {Shapley},
  {Reddy}, {Coil}, {Mobasher}, {Muratov}, {Azadi}, {Leung}, {Sanders},
  {Shivaei}, {Price}, {DeGroot}, \& {Kere{\v{s}}}}]{Freeman2019}
{Freeman}, W.~R., {Siana}, B., {Kriek}, M., {et~al.} 2019, \apj, 873, 102,
  \dodoi{10.3847/1538-4357/ab0655}

\bibitem[{{Fujimoto} {et~al.}(2023{\natexlab{a}}){Fujimoto}, {Wang}, {Weaver},
  {Kokorev}, {Atek}, {Bezanson}, {Labbe}, {Brammer}, {Greene}, {Chemerynska},
  {Dayal}, {de Graaff}, {Furtak}, {Oesch}, {Setton}, {Price}, {Miller},
  {Williams}, {Whitaker}, {Zitrin}, {Cutler}, {Leja}, {Pan}, {Coe}, {van
  Dokkum}, {Feldmann}, {Fudamoto}, {Goulding}, {Khullar}, {Marchesini},
  {Maseda}, {Nanayakkara}, {Nelson}, {Smit}, {Stefanon}, \&
  {Weibel}}]{Fujimoto2023}
{Fujimoto}, S., {Wang}, B., {Weaver}, J., {et~al.} 2023{\natexlab{a}}, arXiv
  e-prints, arXiv:2308.11609, \dodoi{10.48550/arXiv.2308.11609}

\bibitem[{{Fujimoto} {et~al.}(2023{\natexlab{b}}){Fujimoto}, {Bezanson},
  {Labbe}, {Brammer}, {Price}, {Wang}, {Weaver}, {Fudamoto}, {Oesch},
  {Williams}, {Dayal}, {Feldmann}, {Greene}, {Leja}, {Whitaker}, {Zitrin},
  {Cutler}, {Furtak}, {Pan}, {Chemerynska}, {Kokorev}, {Miller}, {Atek}, {van
  Dokkum}, {Juneau}, {Kassin}, {Khullar}, {Marchesini}, {Maseda}, {Nelson},
  {Setton}, \& {Smit}}]{Fujimoto2023:alma}
{Fujimoto}, S., {Bezanson}, R., {Labbe}, I., {et~al.} 2023{\natexlab{b}}, arXiv
  e-prints, arXiv:2309.07834, \dodoi{10.48550/arXiv.2309.07834}

\bibitem[{{Furtak} {et~al.}(2021){Furtak}, {Atek}, {Lehnert}, {Chevallard}, \&
  {Charlot}}]{Furtak2021}
{Furtak}, L.~J., {Atek}, H., {Lehnert}, M.~D., {Chevallard}, J., \& {Charlot},
  S. 2021, \mnras, 501, 1568, \dodoi{10.1093/mnras/staa3760}

\bibitem[{{Furtak} {et~al.}(2023){Furtak}, {Zitrin}, {Weaver}, {Atek},
  {Bezanson}, {Labb{\'e}}, {Whitaker}, {Leja}, {Price}, {Brammer}, {Wang},
  {Marchesini}, {Pan}, {Dayal}, {van Dokkum}, {Feldmann}, {Fujimoto}, {Franx},
  {Khullar}, {Nelson}, \& {Mowla}}]{Furtak2022}
{Furtak}, L.~J., {Zitrin}, A., {Weaver}, J.~R., {et~al.} 2023, \mnras, 523,
  4568, \dodoi{10.1093/mnras/stad1627}

\bibitem[{{Gallazzi} {et~al.}(2005){Gallazzi}, {Charlot}, {Brinchmann},
  {White}, \& {Tremonti}}]{Gallazzi2005}
{Gallazzi}, A., {Charlot}, S., {Brinchmann}, J., {White}, S. D.~M., \&
  {Tremonti}, C.~A. 2005, \mnras, 362, 41,
  \dodoi{10.1111/j.1365-2966.2005.09321.x}

\bibitem[{{Goodman} \& {Weare}(2010)}]{Goodman2010}
{Goodman}, J., \& {Weare}, J. 2010, Communications in Applied Mathematics and
  Computational Science, 5, 65, \dodoi{10.2140/camcos.2010.5.65}

\bibitem[{{Gould} {et~al.}(2023){Gould}, {Brammer}, {Valentino}, {Whitaker},
  {Weaver}, {Lagos}, {Rizzo}, {Franco}, {Hsieh}, {Ilbert}, {Jin}, {Magdis},
  {McCracken}, {Mobasher}, {Shuntov}, {Steinhardt}, {Strait}, \&
  {Toft}}]{Gould2023}
{Gould}, K. M.~L., {Brammer}, G., {Valentino}, F., {et~al.} 2023, \aj, 165,
  248, \dodoi{10.3847/1538-3881/accadc}

\bibitem[{{Goulding} {et~al.}(2023){Goulding}, {Greene}, {Setton}, {Labbe},
  {Bezanson}, {Miller}, {Atek}, {Bogd{\'a}n}, {Brammer}, {Chemerynska},
  {Cutler}, {Dayal}, {Fudamoto}, {Fujimoto}, {Furtak}, {Kokorev}, {Khullar},
  {Leja}, {Marchesini}, {Natarajan}, {Nelson}, {Oesch}, {Pan}, {Papovich},
  {Price}, {van Dokkum}, {Wang}, {Weaver}, {Whitaker}, \&
  {Zitrin}}]{goulding2023}
{Goulding}, A.~D., {Greene}, J.~E., {Setton}, D.~J., {et~al.} 2023, \apjl, 955,
  L24, \dodoi{10.3847/2041-8213/acf7c5}

\bibitem[{{Grogin} {et~al.}(2011){Grogin}, {Kocevski}, {Faber}, {Ferguson},
  {Koekemoer}, {Riess}, {Acquaviva}, {Alexander}, {Almaini}, {Ashby}, {Barden},
  {Bell}, {Bournaud}, {Brown}, {Caputi}, {Casertano}, {Cassata}, {Castellano},
  {Challis}, {Chary}, {Cheung}, {Cirasuolo}, {Conselice}, {Roshan Cooray},
  {Croton}, {Daddi}, {Dahlen}, {Dav{\'e}}, {de Mello}, {Dekel}, {Dickinson},
  {Dolch}, {Donley}, {Dunlop}, {Dutton}, {Elbaz}, {Fazio}, {Filippenko},
  {Finkelstein}, {Fontana}, {Gardner}, {Garnavich}, {Gawiser}, {Giavalisco},
  {Grazian}, {Guo}, {Hathi}, {H{\"a}ussler}, {Hopkins}, {Huang}, {Huang},
  {Jha}, {Kartaltepe}, {Kirshner}, {Koo}, {Lai}, {Lee}, {Li}, {Lotz}, {Lucas},
  {Madau}, {McCarthy}, {McGrath}, {McIntosh}, {McLure}, {Mobasher},
  {Moustakas}, {Mozena}, {Nandra}, {Newman}, {Niemi}, {Noeske}, {Papovich},
  {Pentericci}, {Pope}, {Primack}, {Rajan}, {Ravindranath}, {Reddy}, {Renzini},
  {Rix}, {Robaina}, {Rodney}, {Rosario}, {Rosati}, {Salimbeni}, {Scarlata},
  {Siana}, {Simard}, {Smidt}, {Somerville}, {Spinrad}, {Straughn}, {Strolger},
  {Telford}, {Teplitz}, {Trump}, {van der Wel}, {Villforth}, {Wechsler},
  {Weiner}, {Wiklind}, {Wild}, {Wilson}, {Wuyts}, {Yan}, \&
  {Yun}}]{2011ApJS..197...35G}
{Grogin}, N.~A., {Kocevski}, D.~D., {Faber}, S.~M., {et~al.} 2011, \apjs, 197,
  35, \dodoi{10.1088/0067-0049/197/2/35}

\bibitem[{{Gutkin} {et~al.}(2016){Gutkin}, {Charlot}, \&
  {Bruzual}}]{Gutkin2016}
{Gutkin}, J., {Charlot}, S., \& {Bruzual}, G. 2016, \mnras, 462, 1757,
  \dodoi{10.1093/mnras/stw1716}

\bibitem[{{Harris} {et~al.}(2020){Harris}, {Millman}, {van der Walt},
  {Gommers}, {Virtanen}, {Cournapeau}, {Wieser}, {Taylor}, {Berg}, {Smith},
  {Kern}, {Picus}, {Hoyer}, {van Kerkwijk}, {Brett}, {Haldane}, {del R{\'\i}o},
  {Wiebe}, {Peterson}, {G{\'e}rard-Marchant}, {Sheppard}, {Reddy}, {Weckesser},
  {Abbasi}, {Gohlke}, \& {Oliphant}}]{2020Natur.585..357H}
{Harris}, C.~R., {Millman}, K.~J., {van der Walt}, S.~J., {et~al.} 2020, \nat,
  585, 357, \dodoi{10.1038/s41586-020-2649-2}

\bibitem[{{Hinshaw} {et~al.}(2013){Hinshaw}, {Larson}, {Komatsu}, {Spergel},
  {Bennett}, {Dunkley}, {Nolta}, {Halpern}, {Hill}, {Odegard}, {Page}, {Smith},
  {Weiland}, {Gold}, {Jarosik}, {Kogut}, {Limon}, {Meyer}, {Tucker}, {Wollack},
  \& {Wright}}]{2013ApJS..208...19H}
{Hinshaw}, G., {Larson}, D., {Komatsu}, E., {et~al.} 2013, \apjs, 208, 19,
  \dodoi{10.1088/0067-0049/208/2/19}

\bibitem[{{Hoaglin} {et~al.}(1983){Hoaglin}, {Mosteller}, \&
  {Tukey}}]{Hoaglin1983}
{Hoaglin}, D.~C., {Mosteller}, F., \& {Tukey}, J.~W. 1983, {Understanding
  robust and exploratory data anlysis}

\bibitem[{{Hunter}(2007)}]{2007CSE.....9...90H}
{Hunter}, J.~D. 2007, Computing in Science and Engineering, 9, 90,
  \dodoi{10.1109/MCSE.2007.55}

\bibitem[{{Johnson} {et~al.}(2021){Johnson}, {Leja}, {Conroy}, \&
  {Speagle}}]{Johnson2021}
{Johnson}, B.~D., {Leja}, J., {Conroy}, C., \& {Speagle}, J.~S. 2021, \apjs,
  254, 22, \dodoi{10.3847/1538-4365/abef67}

\bibitem[{{Kannan} {et~al.}(2023){Kannan}, {Springel}, {Hernquist}, {Pakmor},
  {Delgado}, {Hadzhiyska}, {Hern{\'a}ndez-Aguayo}, {Barrera}, {Ferlito},
  {Bose}, {White}, {Frenk}, {Smith}, \& {Garaldi}}]{Kannan2022}
{Kannan}, R., {Springel}, V., {Hernquist}, L., {et~al.} 2023, \mnras, 524,
  2594, \dodoi{10.1093/mnras/stac3743}

\bibitem[{{Kodra} {et~al.}(2023){Kodra}, {Andrews}, {Newman}, {Finkelstein},
  {Fontana}, {Hathi}, {Salvato}, {Wiklind}, {Wuyts}, {Broussard}, {Chartab},
  {Conselice}, {Cooper}, {Dekel}, {Dickinson}, {Ferguson}, {Gawiser}, {Grogin},
  {Iyer}, {Kartaltepe}, {Kassin}, {Koekemoer}, {Koo}, {Lucas}, {Mantha},
  {McIntosh}, {Mobasher}, {Pacifici}, {P{\'e}rez-Gonz{\'a}lez}, \&
  {Santini}}]{2023ApJ...942...36K}
{Kodra}, D., {Andrews}, B.~H., {Newman}, J.~A., {et~al.} 2023, \apj, 942, 36,
  \dodoi{10.3847/1538-4357/ac9f12}

\bibitem[{{Kriek} {et~al.}(2016){Kriek}, {Conroy}, {van Dokkum}, {Shapley},
  {Choi}, {Reddy}, {Siana}, {van de Voort}, {Coil}, \& {Mobasher}}]{Kriek2016}
{Kriek}, M., {Conroy}, C., {van Dokkum}, P.~G., {et~al.} 2016, \nat, 540, 248,
  \dodoi{10.1038/nature20570}

\bibitem[{{Kroupa}(2001)}]{Kroupa2001}
{Kroupa}, P. 2001, \mnras, 322, 231, \dodoi{10.1046/j.1365-8711.2001.04022.x}

\bibitem[{{La Barbera} {et~al.}(2013){La Barbera}, {Ferreras}, {Vazdekis}, {de
  la Rosa}, {de Carvalho}, {Trevisan}, {Falc{\'o}n-Barroso}, \&
  {Ricciardelli}}]{2013MNRAS.433.3017L}
{La Barbera}, F., {Ferreras}, I., {Vazdekis}, A., {et~al.} 2013, \mnras, 433,
  3017, \dodoi{10.1093/mnras/stt943}

\bibitem[{{Labb{\'e}} {et~al.}(2003){Labb{\'e}}, {Franx}, {Rudnick},
  {F{\"o}rster Schreiber}, {Rix}, {Moorwood}, {van Dokkum}, {van der Werf},
  {R{\"o}ttgering}, {van Starkenburg}, {van der Wel}, {Kuijken}, \&
  {Daddi}}]{Labbe2003}
{Labb{\'e}}, I., {Franx}, M., {Rudnick}, G., {et~al.} 2003, \aj, 125, 1107,
  \dodoi{10.1086/346140}

\bibitem[{{Labb{\'e}} {et~al.}(2023){Labb{\'e}}, {van Dokkum}, {Nelson},
  {Bezanson}, {Suess}, {Leja}, {Brammer}, {Whitaker}, {Mathews}, {Stefanon}, \&
  {Wang}}]{Labbe2022}
{Labb{\'e}}, I., {van Dokkum}, P., {Nelson}, E., {et~al.} 2023, \nat, 616, 266,
  \dodoi{10.1038/s41586-023-05786-2}

\bibitem[{{Lagattuta} {et~al.}(2017){Lagattuta}, {Mould}, {Forbes}, {Monson},
  {Pastorello}, \& {Persson}}]{Lagattuta2017}
{Lagattuta}, D.~J., {Mould}, J.~R., {Forbes}, D.~A., {et~al.} 2017, \apj, 846,
  166, \dodoi{10.3847/1538-4357/aa8563}

\bibitem[{{Leistedt} {et~al.}(2023){Leistedt}, {Alsing}, {Peiris}, {Mortlock},
  \& {Leja}}]{Leistedt2023}
{Leistedt}, B., {Alsing}, J., {Peiris}, H., {Mortlock}, D., \& {Leja}, J. 2023,
  \apjs, 264, 23, \dodoi{10.3847/1538-4365/ac9d99}

\bibitem[{{Leitherer} {et~al.}(1999){Leitherer}, {Schaerer}, {Goldader},
  {Delgado}, {Robert}, {Kune}, {de Mello}, {Devost}, \&
  {Heckman}}]{Leitherer1999}
{Leitherer}, C., {Schaerer}, D., {Goldader}, J.~D., {et~al.} 1999, \apjs, 123,
  3, \dodoi{10.1086/313233}

\bibitem[{{Leja} {et~al.}(2019{\natexlab{a}}){Leja}, {Carnall}, {Johnson},
  {Conroy}, \& {Speagle}}]{Leja2019}
{Leja}, J., {Carnall}, A.~C., {Johnson}, B.~D., {Conroy}, C., \& {Speagle},
  J.~S. 2019{\natexlab{a}}, \apj, 876, 3, \dodoi{10.3847/1538-4357/ab133c}

\bibitem[{{Leja} {et~al.}(2017){Leja}, {Johnson}, {Conroy}, {van Dokkum}, \&
  {Byler}}]{Leja2017}
{Leja}, J., {Johnson}, B.~D., {Conroy}, C., {van Dokkum}, P.~G., \& {Byler}, N.
  2017, \apj, 837, 170, \dodoi{10.3847/1538-4357/aa5ffe}

\bibitem[{{Leja} {et~al.}(2020){Leja}, {Speagle}, {Johnson}, {Conroy}, {van
  Dokkum}, \& {Franx}}]{Leja2020}
{Leja}, J., {Speagle}, J.~S., {Johnson}, B.~D., {et~al.} 2020, \apj, 893, 111,
  \dodoi{10.3847/1538-4357/ab7e27}

\bibitem[{{Leja} {et~al.}(2019{\natexlab{b}}){Leja}, {Tacchella}, \&
  {Conroy}}]{Leja2019c}
{Leja}, J., {Tacchella}, S., \& {Conroy}, C. 2019{\natexlab{b}}, \apjl, 880,
  L9, \dodoi{10.3847/2041-8213/ab2f8c}

\bibitem[{{Leja} {et~al.}(2019{\natexlab{c}}){Leja}, {Johnson}, {Conroy}, {van
  Dokkum}, {Speagle}, {Brammer}, {Momcheva}, {Skelton}, {Whitaker}, {Franx}, \&
  {Nelson}}]{Leja2019b}
{Leja}, J., {Johnson}, B.~D., {Conroy}, C., {et~al.} 2019{\natexlab{c}}, \apj,
  877, 140, \dodoi{10.3847/1538-4357/ab1d5a10.48550/arXiv.1812.05608}

\bibitem[{{Lotz} {et~al.}(2017){Lotz}, {Koekemoer}, {Coe}, {Grogin}, {Capak},
  {Mack}, {Anderson}, {Avila}, {Barker}, {Borncamp}, {Brammer}, {Durbin},
  {Gunning}, {Hilbert}, {Jenkner}, {Khandrika}, {Levay}, {Lucas}, {MacKenty},
  {Ogaz}, {Porterfield}, {Reid}, {Robberto}, {Royle}, {Smith},
  {Storrie-Lombardi}, {Sunnquist}, {Surace}, {Taylor}, {Williams}, {Bullock},
  {Dickinson}, {Finkelstein}, {Natarajan}, {Richard}, {Robertson}, {Tumlinson},
  {Zitrin}, {Flanagan}, {Sembach}, {Soifer}, \& {Mountain}}]{Lotz2017}
{Lotz}, J.~M., {Koekemoer}, A., {Coe}, D., {et~al.} 2017, \apj, 837, 97,
  \dodoi{10.3847/1538-4357/837/1/97}

\bibitem[{{Lyubenova} {et~al.}(2016){Lyubenova}, {Mart{\'\i}n-Navarro}, {van de
  Ven}, {Falc{\'o}n-Barroso}, {Galbany}, {Gallazzi}, {Garc{\'\i}a-Benito},
  {Gonz{\'a}lez Delgado}, {Husemann}, {La Barbera}, {Marino}, {Mast},
  {Mendez-Abreu}, {Peletier}, {S{\'a}nchez-Bl{\'a}zquez}, {S{\'a}nchez},
  {Trager}, {van den Bosch}, {Vazdekis}, {Walcher}, {Zhu}, {Zibetti},
  {Ziegler}, {Bland-Hawthorn}, \& {CALIFA Collaboration}}]{Lyubenova2016}
{Lyubenova}, M., {Mart{\'\i}n-Navarro}, I., {van de Ven}, G., {et~al.} 2016,
  \mnras, 463, 3220, \dodoi{10.1093/mnras/stw2434}

\bibitem[{{Ma} {et~al.}(2016){Ma}, {Hopkins}, {Faucher-Gigu{\`e}re}, {Zolman},
  {Muratov}, {Kere{\v{s}}}, \& {Quataert}}]{Ma2016}
{Ma}, X., {Hopkins}, P.~F., {Faucher-Gigu{\`e}re}, C.-A., {et~al.} 2016,
  \mnras, 456, 2140, \dodoi{10.1093/mnras/stv2659}

\bibitem[{{Madau}(1995)}]{Madau1995}
{Madau}, P. 1995, \apj, 441, 18, \dodoi{10.1086/175332}

\bibitem[{{Marchesini} {et~al.}(2023){Marchesini}, {Brammer}, {Morishita},
  {Bergamini}, {Wang}, {Bradac}, {Roberts-Borsani}, {Strait}, {Treu},
  {Fontana}, {Jones}, {Santini}, {Vulcani}, {Acebron}, {Calabr{\`o}},
  {Castellano}, {Glazebrook}, {Grillo}, {Mercurio}, {Nanayakkara}, {Rosati},
  {Tubthong}, \& {Vanzella}}]{Marchesini2023}
{Marchesini}, D., {Brammer}, G., {Morishita}, T., {et~al.} 2023, \apjl, 942,
  L25, \dodoi{10.3847/2041-8213/acaaac}

\bibitem[{{Mathews} {et~al.}(2023){Mathews}, {Leja}, {Speagle}, {Johnson},
  {Gibson}, {Nelson}, {Suess}, {Tacchella}, {Whitaker}, \&
  {Wang}}]{Mathews2023}
{Mathews}, E.~P., {Leja}, J., {Speagle}, J.~S., {et~al.} 2023, \apj, 954, 132,
  \dodoi{10.3847/1538-4357/ace720}

\bibitem[{{Mauerhofer} \& {Dayal}(2023)}]{Mauerhofer2023}
{Mauerhofer}, V., \& {Dayal}, P. 2023, \mnras, 526, 2196,
  \dodoi{10.1093/mnras/stad2734}

\bibitem[{{McKinney} {et~al.}(2023){McKinney}, {Finnerty}, {Casey}, {Franco},
  {Long}, {Fujimoto}, {Zavala}, {Cooper}, {Akins}, {Pope}, {Armus}, {Soifer},
  {Larson}, {Matthews}, {Melbourne}, \& {Cushing}}]{McKinney2023}
{McKinney}, J., {Finnerty}, L., {Casey}, C.~M., {et~al.} 2023, \apjl, 946, L39,
  \dodoi{10.3847/2041-8213/acc322}

\bibitem[{{Meneghetti} {et~al.}(2017){Meneghetti}, {Natarajan}, {Coe},
  {Contini}, {De Lucia}, {Giocoli}, {Acebron}, {Borgani}, {Bradac}, {Diego},
  {Hoag}, {Ishigaki}, {Johnson}, {Jullo}, {Kawamata}, {Lam}, {Limousin},
  {Liesenborgs}, {Oguri}, {Sebesta}, {Sharon}, {Williams}, \&
  {Zitrin}}]{Meneghetti2017}
{Meneghetti}, M., {Natarajan}, P., {Coe}, D., {et~al.} 2017, \mnras, 472, 3177,
  \dodoi{10.1093/mnras/stx2064}

\bibitem[{{Mirocha} \& {Furlanetto}(2023)}]{Mirocha2023}
{Mirocha}, J., \& {Furlanetto}, S.~R. 2023, \mnras, 519, 843,
  \dodoi{10.1093/mnras/stac3578}

\bibitem[{{Mitchell} {et~al.}(2013){Mitchell}, {Lacey}, {Baugh}, \&
  {Cole}}]{Mitchell2013}
{Mitchell}, P.~D., {Lacey}, C.~G., {Baugh}, C.~M., \& {Cole}, S. 2013, \mnras,
  435, 87, \dodoi{10.1093/mnras/stt1280}

\bibitem[{{Naidu} {et~al.}(2022){Naidu}, {Oesch}, {Setton}, {Matthee},
  {Conroy}, {Johnson}, {Weaver}, {Bouwens}, {Brammer}, {Dayal}, {Illingworth},
  {Barrufet}, {Belli}, {Bezanson}, {Bose}, {Heintz}, {Leja}, {Leonova},
  {Marques-Chaves}, {Stefanon}, {Toft}, {van der Wel}, {van Dokkum}, {Weibel},
  \& {Whitaker}}]{Naidu2022}
{Naidu}, R.~P., {Oesch}, P.~A., {Setton}, D.~J., {et~al.} 2022, arXiv e-prints,
  arXiv:2208.02794.
\newblock \doarXiv{2208.02794}

\bibitem[{{Newman} \& {Gruen}(2022)}]{Newman2022}
{Newman}, J.~A., \& {Gruen}, D. 2022, \araa, 60, 363,
  \dodoi{10.1146/annurev-astro-032122-014611}

\bibitem[{{Noll} {et~al.}(2009){Noll}, {Burgarella}, {Giovannoli}, {Buat},
  {Marcillac}, \& {Mu{\~n}oz-Mateos}}]{Noll2009}
{Noll}, S., {Burgarella}, D., {Giovannoli}, E., {et~al.} 2009, \aap, 507, 1793,
  \dodoi{10.1051/0004-6361/200912497}

\bibitem[{{Onodera} {et~al.}(2015){Onodera}, {Carollo}, {Renzini},
  {Cappellari}, {Mancini}, {Arimoto}, {Daddi}, {Gobat}, {Strazzullo},
  {Tacchella}, \& {Yamada}}]{Onodera2015}
{Onodera}, M., {Carollo}, C.~M., {Renzini}, A., {et~al.} 2015, \apj, 808, 161,
  \dodoi{10.1088/0004-637X/808/2/161}

\bibitem[{{Pacifici} {et~al.}(2023){Pacifici}, {Iyer}, {Mobasher}, {da Cunha},
  {Acquaviva}, {Burgarella}, {Calistro Rivera}, {Carnall}, {Chang}, {Chartab},
  {Cooke}, {Fairhurst}, {Kartaltepe}, {Leja}, {Ma{\l}ek}, {Salmon}, {Torelli},
  {Vidal-Garc{\'\i}a}, {Boquien}, {Brammer}, {Brown}, {Capak}, {Chevallard},
  {Circosta}, {Croton}, {Davidzon}, {Dickinson}, {Duncan}, {Faber}, {Ferguson},
  {Fontana}, {Guo}, {Haeussler}, {Hemmati}, {Jafariyazani}, {Kassin}, {Larson},
  {Lee}, {Mantha}, {Marchi}, {Nayyeri}, {Newman}, {Pandya}, {Pforr}, {Reddy},
  {Sanders}, {Shah}, {Shahidi}, {Stevans}, {Triani}, {Tyler}, {Vanderhoof}, {de
  la Vega}, {Wang}, \& {Weston}}]{Pacifici2022}
{Pacifici}, C., {Iyer}, K.~G., {Mobasher}, B., {et~al.} 2023, \apj, 944, 141,
  \dodoi{10.3847/1538-4357/acacff}

\bibitem[{{Reddy} {et~al.}(2023){Reddy}, {Topping}, {Sanders}, {Shapley}, \&
  {Brammer}}]{Reddy2023}
{Reddy}, N.~A., {Topping}, M.~W., {Sanders}, R.~L., {Shapley}, A.~E., \&
  {Brammer}, G. 2023, \apj, 952, 167, \dodoi{10.3847/1538-4357/acd754}

\bibitem[{{Richard} {et~al.}(2021){Richard}, {Claeyssens}, {Lagattuta},
  {Guaita}, {Bauer}, {Pello}, {Carton}, {Bacon}, {Soucail}, {Lyon}, {Kneib},
  {Mahler}, {Cl{\'e}ment}, {Mercier}, {Variu}, {Tamone}, {Ebeling}, {Schmidt},
  {Nanayakkara}, {Maseda}, {Weilbacher}, {Bouch{\'e}}, {Bouwens}, {Wisotzki},
  {de la Vieuville}, {Martinez}, \& {Patr{\'\i}cio}}]{Richard2021}
{Richard}, J., {Claeyssens}, A., {Lagattuta}, D., {et~al.} 2021, \aap, 646,
  A83, \dodoi{10.1051/0004-6361/202039462}

\bibitem[{{Rieke} {et~al.}(2023){Rieke}, {Kelly}, {Misselt}, {Stansberry},
  {Boyer}, {Beatty}, {Egami}, {Florian}, {Greene}, {Hainline}, {Leisenring},
  {Roellig}, {Schlawin}, {Sun}, {Tinnin}, {Williams}, {Willmer}, {Wilson},
  {Clark}, {Rohrbach}, {Brooks}, {Canipe}, {Correnti}, {DiFelice}, {Gennaro},
  {Girard}, {Hartig}, {Hilbert}, {Koekemoer}, {Nikolov}, {Pirzkal}, {Rest},
  {Robberto}, {Sunnquist}, {Telfer}, {Wu}, {Ferry}, {Lewis}, {Baum},
  {Beichman}, {Doyon}, {Dressler}, {Eisenstein}, {Ferrarese}, {Hodapp},
  {Horner}, {Jaffe}, {Johnstone}, {Krist}, {Martin}, {McCarthy}, {Meyer},
  {Rieke}, {Trauger}, \& {Young}}]{Rieke2023}
{Rieke}, M.~J., {Kelly}, D.~M., {Misselt}, K., {et~al.} 2023, \pasp, 135,
  028001, \dodoi{10.1088/1538-3873/acac53}

\bibitem[{{Rigby} {et~al.}(2023){Rigby}, {Perrin}, {McElwain}, {Kimble},
  {Friedman}, {Lallo}, {Doyon}, {Feinberg}, {Ferruit}, {Glasse}, \&
  et~al.}]{Rigby2023}
{Rigby}, J., {Perrin}, M., {McElwain}, M., {et~al.} 2023, \pasp, 135, 048001,
  \dodoi{10.1088/1538-3873/acb293}

\bibitem[{{Salpeter}(1955)}]{Salpeter1955}
{Salpeter}, E.~E. 1955, \apj, 121, 161, \dodoi{10.1086/145971}

\bibitem[{{S{\'a}nchez-Bl{\'a}zquez} {et~al.}(2006){S{\'a}nchez-Bl{\'a}zquez},
  {Peletier}, {Jim{\'e}nez-Vicente}, {Cardiel}, {Cenarro},
  {Falc{\'o}n-Barroso}, {Gorgas}, {Selam}, \& {Vazdekis}}]{2006MNRAS.371..703S}
{S{\'a}nchez-Bl{\'a}zquez}, P., {Peletier}, R.~F., {Jim{\'e}nez-Vicente}, J.,
  {et~al.} 2006, \mnras, 371, 703, \dodoi{10.1111/j.1365-2966.2006.10699.x}

\bibitem[{{Schaerer} \& {de Barros}(2009)}]{Schaerer2009}
{Schaerer}, D., \& {de Barros}, S. 2009, \aap, 502, 423,
  \dodoi{10.1051/0004-6361/200911781}

\bibitem[{{Skilling}(2004)}]{Skilling2004}
{Skilling}, J. 2004, in American Institute of Physics Conference Series, Vol.
  735, Bayesian Inference and Maximum Entropy Methods in Science and
  Engineering: 24th International Workshop on Bayesian Inference and Maximum
  Entropy Methods in Science and Engineering, ed. R.~{Fischer}, R.~{Preuss}, \&
  U.~V. {Toussaint}, 395--405, \dodoi{10.1063/1.1835238}

\bibitem[{{Smit} {et~al.}(2014){Smit}, {Bouwens}, {Labb{\'e}}, {Zheng},
  {Bradley}, {Donahue}, {Lemze}, {Moustakas}, {Umetsu}, {Zitrin}, {Coe},
  {Postman}, {Gonzalez}, {Bartelmann}, {Ben{\'\i}tez}, {Broadhurst}, {Ford},
  {Grillo}, {Infante}, {Jimenez-Teja}, {Jouvel}, {Kelson}, {Lahav}, {Maoz},
  {Medezinski}, {Melchior}, {Meneghetti}, {Merten}, {Molino}, {Moustakas},
  {Nonino}, {Rosati}, \& {Seitz}}]{Smit2014}
{Smit}, R., {Bouwens}, R.~J., {Labb{\'e}}, I., {et~al.} 2014, \apj, 784, 58,
  \dodoi{10.1088/0004-637X/784/1/58}

\bibitem[{{Speagle}(2020)}]{Speagle2020}
{Speagle}, J.~S. 2020, \mnras, 493, 3132, \dodoi{10.1093/mnras/staa278}

\bibitem[{{Spiniello} {et~al.}(2014){Spiniello}, {Trager}, {Koopmans}, \&
  {Conroy}}]{Spiniello2014}
{Spiniello}, C., {Trager}, S., {Koopmans}, L. V.~E., \& {Conroy}, C. 2014,
  \mnras, 438, 1483, \dodoi{10.1093/mnras/stt2282}

\bibitem[{{Stanway} \& {Eldridge}(2018)}]{Stanway2018}
{Stanway}, E.~R., \& {Eldridge}, J.~J. 2018, \mnras, 479, 75,
  \dodoi{10.1093/mnras/sty1353}

\bibitem[{{Stark} {et~al.}(2013){Stark}, {Schenker}, {Ellis}, {Robertson},
  {McLure}, \& {Dunlop}}]{Stark2013}
{Stark}, D.~P., {Schenker}, M.~A., {Ellis}, R., {et~al.} 2013, \apj, 763, 129,
  \dodoi{10.1088/0004-637X/763/2/129}

\bibitem[{{Steidel} {et~al.}(1996){Steidel}, {Giavalisco}, {Pettini},
  {Dickinson}, \& {Adelberger}}]{Steidel1996}
{Steidel}, C.~C., {Giavalisco}, M., {Pettini}, M., {Dickinson}, M., \&
  {Adelberger}, K.~L. 1996, \apjl, 462, L17, \dodoi{10.1086/310029}

\bibitem[{{Steidel} {et~al.}(2016){Steidel}, {Strom}, {Pettini}, {Rudie},
  {Reddy}, \& {Trainor}}]{Steidel2016}
{Steidel}, C.~C., {Strom}, A.~L., {Pettini}, M., {et~al.} 2016, \apj, 826, 159,
  \dodoi{10.3847/0004-637X/826/2/159}

\bibitem[{{Steinhardt} {et~al.}(2023){Steinhardt}, {Kokorev}, {Rusakov},
  {Garcia}, \& {Sneppen}}]{Steinhardt2022}
{Steinhardt}, C.~L., {Kokorev}, V., {Rusakov}, V., {Garcia}, E., \& {Sneppen},
  A. 2023, \apjl, 951, L40, \dodoi{10.3847/2041-8213/acdef6}

\bibitem[{{Steinhardt} {et~al.}(2020){Steinhardt}, {Jauzac}, {Acebron}, {Atek},
  {Capak}, {Davidzon}, {Eckert}, {Harvey}, {Koekemoer}, {Lagos}, {Mahler},
  {Montes}, {Niemiec}, {Nonino}, {Oesch}, {Richard}, {Rodney}, {Schaller},
  {Sharon}, {Strolger}, {Allingham}, {Amara}, {Bah{\'e}}, {B{\oe}hm}, {Bose},
  {Bouwens}, {Bradley}, {Brammer}, {Broadhurst}, {Ca{\~n}as}, {Cen},
  {Cl{\'e}ment}, {Clowe}, {Coe}, {Connor}, {Darvish}, {Diego}, {Ebeling},
  {Edge}, {Egami}, {Ettori}, {Faisst}, {Frye}, {Furtak}, {G{\'o}mez-Guijarro},
  {Remolina Gonz{\'a}lez}, {Gonzalez}, {Graur}, {Gruen}, {Harvey}, {Hensley},
  {Hovis-Afflerbach}, {Jablonka}, {Jha}, {Jullo}, {Kneib}, {Kokorev},
  {Lagattuta}, {Limousin}, {von der Linden}, {Linzer}, {Lopez}, {Magdis},
  {Massey}, {Masters}, {Maturi}, {McCully}, {McGee}, {Meneghetti}, {Mobasher},
  {Moustakas}, {Murphy}, {Natarajan}, {Neyrinck}, {O'Connor}, {Oguri}, {Pagul},
  {Rhodes}, {Rich}, {Robertson}, {Sereno}, {Shan}, {Smith}, {Sneppen},
  {Squires}, {Tam}, {Tchernin}, {Toft}, {Umetsu}, {Weaver}, {van Weeren},
  {Williams}, {Wilson}, {Yan}, \& {Zitrin}}]{Steinhardt2020}
{Steinhardt}, C.~L., {Jauzac}, M., {Acebron}, A., {et~al.} 2020, \apjs, 247,
  64, \dodoi{10.3847/1538-4365/ab75ed}

\bibitem[{{Strom} {et~al.}(2018){Strom}, {Steidel}, {Rudie}, {Trainor}, \&
  {Pettini}}]{Strom2018}
{Strom}, A.~L., {Steidel}, C.~C., {Rudie}, G.~C., {Trainor}, R.~F., \&
  {Pettini}, M. 2018, \apj, 868, 117, \dodoi{10.3847/1538-4357/aae1a5}

\bibitem[{{Tacchella} {et~al.}(2022){Tacchella}, {Conroy}, {Faber}, {Johnson},
  {Leja}, {Barro}, {Cunningham}, {Deason}, {Guhathakurta}, {Guo}, {Hernquist},
  {Koo}, {McKinnon}, {Rockosi}, {Speagle}, {van Dokkum}, \&
  {Yesuf}}]{Tacchella2022}
{Tacchella}, S., {Conroy}, C., {Faber}, S.~M., {et~al.} 2022, \apj, 926, 134,
  \dodoi{10.3847/1538-4357/ac449b}

\bibitem[{{Thomas} {et~al.}(2005){Thomas}, {Maraston}, {Bender}, \& {Mendes de
  Oliveira}}]{Thomas2005}
{Thomas}, D., {Maraston}, C., {Bender}, R., \& {Mendes de Oliveira}, C. 2005,
  \apj, 621, 673, \dodoi{10.1086/426932}

\bibitem[{{Thorne} {et~al.}(2022){Thorne}, {Robotham}, {Bellstedt}, {Davies},
  {Cook}, {Cortese}, {Holwerda}, {Phillipps}, \& {Siudek}}]{Thorne2022}
{Thorne}, J.~E., {Robotham}, A. S.~G., {Bellstedt}, S., {et~al.} 2022, \mnras,
  517, 6035, \dodoi{10.1093/mnras/stac3082}

\bibitem[{{Topping} {et~al.}(2022){Topping}, {Stark}, {Endsley}, {Bouwens},
  {Schouws}, {Smit}, {Stefanon}, {Inami}, {Bowler}, {Oesch}, {Gonzalez},
  {Dayal}, {da Cunha}, {Algera}, {van der Werf}, {Pallottini}, {Barrufet},
  {Schneider}, {De Looze}, {Sommovigo}, {Whitler}, {Graziani}, {Fudamoto}, \&
  {Ferrara}}]{Topping2022}
{Topping}, M.~W., {Stark}, D.~P., {Endsley}, R., {et~al.} 2022, \mnras, 516,
  975, \dodoi{10.1093/mnras/stac2291}

\bibitem[{{Treu} {et~al.}(2015){Treu}, {Schmidt}, {Brammer}, {Vulcani}, {Wang},
  {Brada{\v{c}}}, {Dijkstra}, {Dressler}, {Fontana}, {Gavazzi}, {Henry},
  {Hoag}, {Huang}, {Jones}, {Kelly}, {Malkan}, {Mason}, {Pentericci},
  {Poggianti}, {Stiavelli}, {Trenti}, \& {von der Linden}}]{Treu2015}
{Treu}, T., {Schmidt}, K.~B., {Brammer}, G.~B., {et~al.} 2015, \apj, 812, 114,
  \dodoi{10.1088/0004-637X/812/2/114}

\bibitem[{{Treu} {et~al.}(2022){Treu}, {Roberts-Borsani}, {Bradac}, {Brammer},
  {Fontana}, {Henry}, {Mason}, {Morishita}, {Pentericci}, {Wang}, {Acebron},
  {Bagley}, {Bergamini}, {Belfiori}, {Bonchi}, {Boyett}, {Boutsia},
  {Calabr{\'o}}, {Caminha}, {Castellano}, {Dressler}, {Glazebrook}, {Grillo},
  {Jacobs}, {Jones}, {Kelly}, {Leethochawalit}, {Malkan}, {Marchesini},
  {Mascia}, {Mercurio}, {Merlin}, {Nanayakkara}, {Nonino}, {Paris},
  {Poggianti}, {Rosati}, {Santini}, {Scarlata}, {Shipley}, {Strait}, {Trenti},
  {Tubthong}, {Vanzella}, {Vulcani}, \& {Yang}}]{Treu2022}
{Treu}, T., {Roberts-Borsani}, G., {Bradac}, M., {et~al.} 2022, \apj, 935, 110,
  \dodoi{10.3847/1538-4357/ac8158}

\bibitem[{{van Dokkum} {et~al.}(2017){van Dokkum}, {Conroy}, {Villaume},
  {Brodie}, \& {Romanowsky}}]{vanDokkum2017}
{van Dokkum}, P., {Conroy}, C., {Villaume}, A., {Brodie}, J., \& {Romanowsky},
  A.~J. 2017, \apj, 841, 68, \dodoi{10.3847/1538-4357/aa7135}

\bibitem[{{Vazdekis} {et~al.}(2015){Vazdekis}, {Coelho}, {Cassisi},
  {Ricciardelli}, {Falc{\'o}n-Barroso}, {S{\'a}nchez-Bl{\'a}zquez}, {La
  Barbera}, {Beasley}, \& {Pietrinferni}}]{Vazdekis2015}
{Vazdekis}, A., {Coelho}, P., {Cassisi}, S., {et~al.} 2015, \mnras, 449, 1177,
  \dodoi{10.1093/mnras/stv151}

\bibitem[{{Virtanen} {et~al.}(2020){Virtanen}, {Gommers}, {Oliphant},
  {Haberland}, {Reddy}, {Cournapeau}, {Burovski}, {Peterson}, {Weckesser},
  {Bright}, {van der Walt}, {Brett}, {Wilson}, {Millman}, {Mayorov}, {Nelson},
  {Jones}, {Kern}, {Larson}, {Carey}, {Polat}, {Feng}, {Moore}, {VanderPlas},
  {Laxalde}, {Perktold}, {Cimrman}, {Henriksen}, {Quintero}, {Harris},
  {Archibald}, {Ribeiro}, {Pedregosa}, {van Mulbregt}, \& {SciPy 1. 0
  Contributors}}]{2020NatMe..17..261V}
{Virtanen}, P., {Gommers}, R., {Oliphant}, T.~E., {et~al.} 2020, NatMe, 17,
  261, \dodoi{10.1038/s41592-019-0686-2}

\bibitem[{{Walcher} {et~al.}(2011){Walcher}, {Groves}, {Budav{\'a}ri}, \&
  {Dale}}]{Walcher2011}
{Walcher}, J., {Groves}, B., {Budav{\'a}ri}, T., \& {Dale}, D. 2011, \apss,
  331, 1, \dodoi{10.1007/s10509-010-0458-z}

\bibitem[{{Wang} {et~al.}(2023{\natexlab{a}}){Wang}, {Leja}, {Villar}, \&
  {Speagle}}]{Wang2023b}
{Wang}, B., {Leja}, J., {Villar}, V.~A., \& {Speagle}, J.~S.
  2023{\natexlab{a}}, \apjl, 952, L10, \dodoi{10.3847/2041-8213/ace361}

\bibitem[{{Wang} {et~al.}(2023{\natexlab{b}}){Wang}, {Fujimoto}, {Labb{\'e}},
  {Furtak}, {Miller}, {Setton}, {Zitrin}, {Atek}, {Bezanson}, {Brammer},
  {Leja}, {Oesch}, {Price}, {Chemerynska}, {Cutler}, {Dayal}, {van Dokkum},
  {Goulding}, {Greene}, {Fudamoto}, {Khullar}, {Kokorev}, {Marchesini}, {Pan},
  {Weaver}, {Whitaker}, \& {Williams}}]{wang2023:z12}
{Wang}, B., {Fujimoto}, S., {Labb{\'e}}, I., {et~al.} 2023{\natexlab{b}},
  \apjl, 957, L34, \dodoi{10.3847/2041-8213/acfe07}

\bibitem[{{Wang} {et~al.}(2023{\natexlab{c}}){Wang}, {Leja}, {Bezanson},
  {Johnson}, {Khullar}, {Labb{\'e}}, {Price}, {Weaver}, \&
  {Whitaker}}]{Wang2023}
{Wang}, B., {Leja}, J., {Bezanson}, R., {et~al.} 2023{\natexlab{c}}, \apjl,
  944, L58, \dodoi{10.3847/2041-8213/acba99}

\bibitem[{{Wang} {et~al.}(2024){Wang}, {Leja}, {Atek}, {Labb{\'e}}, {Li},
  {Bezanson}, {Brammer}, {Cutler}, {Dayal}, {Furtak}, {Greene}, {Kokorev},
  {Pan}, {Price}, {Suess}, {Weaver}, {Whitaker}, \& {Williams}}]{Wang2023:sys}
{Wang}, B., {Leja}, J., {Atek}, H., {et~al.} 2024, \apj, 963, 74,
  \dodoi{10.3847/1538-4357/ad187c}

\bibitem[{{Weaver} {et~al.}(2024){Weaver}, {Cutler}, {Pan}, {Whitaker},
  {Labb{\'e}}, {Price}, {Bezanson}, {Brammer}, {Marchesini}, {Leja}, {Wang},
  {Furtak}, {Zitrin}, {Atek}, {Chemerynska}, {Coe}, {Dayal}, {van Dokkum},
  {Feldmann}, {F{\"o}rster Schreiber}, {Franx}, {Fujimoto}, {Fudamoto},
  {Glazebrook}, {de Graaff}, {Greene}, {Juneau}, {Kassin}, {Kriek}, {Khullar},
  {Maseda}, {Mowla}, {Muzzin}, {Nanayakkara}, {Nelson}, {Oesch}, {Pacifici},
  {Papovich}, {Setton}, {Shapley}, {Shipley}, {Smit}, {Stefanon}, {Taylor},
  {Weibel}, \& {Williams}}]{Weaver2023}
{Weaver}, J.~R., {Cutler}, S.~E., {Pan}, R., {et~al.} 2024, \apjs, 270, 7,
  \dodoi{10.3847/1538-4365/ad07e0}

\bibitem[{{Webb} {et~al.}(2020){Webb}, {Balogh}, {Leja}, {van der Burg},
  {Rudnick}, {Muzzin}, {Boak}, {Cerulo}, {Gilbank}, {Lidman}, {Old},
  {Pintos-Castro}, {McGee}, {Shipley}, {Biviano}, {Chan}, {Cooper}, {De Lucia},
  {Demarco}, {Forrest}, {Jablonka}, {Kukstas}, {McCarthy}, {McNab}, {Nantais},
  {Noble}, {Poggianti}, {Reeves}, {Vulcani}, {Wilson}, {Yee}, \&
  {Zaritsky}}]{Webb2020}
{Webb}, K., {Balogh}, M.~L., {Leja}, J., {et~al.} 2020, \mnras, 498, 5317,
  \dodoi{10.1093/mnras/staa2752}

\bibitem[{{Whitler} {et~al.}(2023){Whitler}, {Stark}, {Endsley}, {Leja},
  {Charlot}, \& {Chevallard}}]{Whitler2022}
{Whitler}, L., {Stark}, D.~P., {Endsley}, R., {et~al.} 2023, \mnras, 519, 5859,
  \dodoi{10.1093/mnras/stad004}

\bibitem[{{Williams} {et~al.}(2009){Williams}, {Quadri}, {Franx}, {van Dokkum},
  \& {Labb{\'e}}}]{Williams2009}
{Williams}, R.~J., {Quadri}, R.~F., {Franx}, M., {van Dokkum}, P., \&
  {Labb{\'e}}, I. 2009, \apj, 691, 1879,
  \dodoi{10.1088/0004-637X/691/2/187910.48550/arXiv.0806.0625}

\bibitem[{{Yung} {et~al.}(2024){Yung}, {Somerville}, {Finkelstein}, {Wilkins},
  \& {Gardner}}]{Yung2023}
{Yung}, L.~Y.~A., {Somerville}, R.~S., {Finkelstein}, S.~L., {Wilkins}, S.~M.,
  \& {Gardner}, J.~P. 2024, \mnras, 527, 5929, \dodoi{10.1093/mnras/stad3484}

\bibitem[{{Zavala} {et~al.}(2023){Zavala}, {Buat}, {Casey}, {Finkelstein},
  {Burgarella}, {Bagley}, {Ciesla}, {Daddi}, {Dickinson}, {Ferguson}, {Franco},
  {Jim{\'e}nez-Andrade}, {Kartaltepe}, {Koekemoer}, {Le Bail}, {Murphy},
  {Papovich}, {Tacchella}, {Wilkins}, {Aretxaga}, {Behroozi}, {Champagne},
  {Fontana}, {Giavalisco}, {Grazian}, {Grogin}, {Kewley}, {Kocevski},
  {Kirkpatrick}, {Lotz}, {Pentericci}, {P{\'e}rez-Gonz{\'a}lez}, {Pirzkal},
  {Ravindranath}, {Somerville}, {Trump}, {Yang}, {Aaron Yung}, {Almaini},
  {Amor{\'\i}n}, {Annunziatella}, {Haro}, {Backhaus}, {Barro}, {Bell},
  {Bhatawdekar}, {Bisigello}, {Buitrago}, {Calabr{\`o}}, {Castellano},
  {Ch{\'a}vez Ortiz}, {Chworowsky}, {Cleri}, {Cohen}, {Cole}, {Cooke},
  {Cooper}, {Cooray}, {Costantin}, {Cox}, {Croton}, {Dav{\'e}}, {de La Vega},
  {Dekel}, {Elbaz}, {Estrada-Carpenter}, {Fern{\'a}ndez}, {Finkelstein},
  {Freundlich}, {Fujimoto}, {Garc{\'\i}a-Argum{\'a}nez}, {Gardner}, {Gawiser},
  {G{\'o}mez-Guijarro}, {Guo}, {Hamilton}, {Hathi}, {Holwerda}, {Hirschmann},
  {Huertas-Company}, {Hutchison}, {Iyer}, {Jaskot}, {Jha}, {Jogee}, {Juneau},
  {Jung}, {Kassin}, {Kurczynski}, {Larson}, {Leung}, {Long}, {Lucas},
  {Magnelli}, {Mantha}, {Matharu}, {McGrath}, {McIntosh}, {Medrano}, {Merlin},
  {Mobasher}, {Morales}, {Newman}, {Nicholls}, {Pandya}, {Rafelski}, {Ronayne},
  {Rose}, {Ryan}, {Santini}, {Seill{\'e}}, {Shah}, {Shen}, {Simons}, {Snyder},
  {Stanway}, {Straughn}, {Teplitz}, {Vanderhoof}, {Vega-Ferrero}, {Wang},
  {Weiner}, {Willmer}, {Wuyts}, \& {Ceers Team}}]{Zavala2023}
{Zavala}, J.~A., {Buat}, V., {Casey}, C.~M., {et~al.} 2023, \apjl, 943, L9,
  \dodoi{10.3847/2041-8213/acacfe}

\bibitem[{{Zitrin} {et~al.}(2015){Zitrin}, {Fabris}, {Merten}, {Melchior},
  {Meneghetti}, {Koekemoer}, {Coe}, {Maturi}, {Bartelmann}, {Postman},
  {Umetsu}, {Seidel}, {Sendra}, {Broadhurst}, {Balestra}, {Biviano}, {Grillo},
  {Mercurio}, {Nonino}, {Rosati}, {Bradley}, {Carrasco}, {Donahue}, {Ford},
  {Frye}, \& {Moustakas}}]{Zitrin2015}
{Zitrin}, A., {Fabris}, A., {Merten}, J., {et~al.} 2015, \apj, 801, 44,
  \dodoi{10.1088/0004-637X/801/1/44}

\end{thebibliography}

\end{document}